\documentclass[twocolumn,reprint,amsmath,amssymb,aps,prab,floatfix,longbibliography,]{revtex4-1}

\usepackage{graphicx}
\usepackage{dcolumn}
\usepackage{bm}
\usepackage{titlesec}
\usepackage{capt-of} 
\usepackage{upgreek}
\usepackage{xcolor}
\usepackage{natbib}
\usepackage{indentfirst}
\titlespacing*{\section}{0pt}{1.2\baselineskip}{1\baselineskip}

\begin{document}


\title{Stability Enhancement of a Self-Amplified Spontaneous Emission Free-electron Laser with Bunching Containment}

\author{Huaiqian Yi \textsuperscript{1}}
\author{Xiaofan Wang \textsuperscript{1}}
\email{wangxf@mail.iasf.ac.cn}
\author{Li Zeng \textsuperscript{1}}
\author{Yifan Liang \textsuperscript{1}}
\author{Weiqing Zhang\textsuperscript{2}}
\email{weiqingzhang@dicp.ac.cn}
\affiliation{
		\textsuperscript{1}Institute of Advanced Science Facilities, Shenzhen 518107, China\\
        \textsuperscript{2}Dalian Institute of Chemical Physics, Chinese Academy of Sciences, Dalian 116023, China}
        
\date{\today}

\begin{abstract}

The self-amplified spontaneous emission (SASE) mechanism, the fundamental operating principle of numerous free-electron laser (FEL) facilities, is driven by electron beam shot noise and leads to significant fluctuations in the output pulse energy. This study presents a robust method for improving pulse energy stability by incorporating a dispersion element that introduces longitudinal dispersion into the electron beam during the exponential growth phase of the SASE process. At this phase, the density modulation of the electron beam, characterized by the bunching factor, undergoes large fluctuations, resulting in substantial variations in the emitted radiation power. The introduction of longitudinal dispersion allows for controlled manipulation of the bunching distribution, suppressing fluctuations and enhancing pulse energy stability. The stabilization mechanism is explained in this paper, and its impact on the radiation properties is analyzed for both the standard SASE scheme and advanced lasing setups, such as a two-stage lasing process for two-color pulse generation, with the initial stage operating in SASE mode.

\end{abstract}

\maketitle


  
    



\section{INTRODUCTION}

X-rays continue to revolutionize the understanding of matter and drive the development of new scientific and technological advancements. High-gain free-electron laser (FEL) amplifiers hold great potential as sources of high-power, coherent, and tunable radiation in the X-ray region of the electromagnetic spectrum. The self-amplified spontaneous emission (SASE) process generates coherent, ultra-bright X-ray beams by amplifying the spontaneous radiation of relativistic electron beams passing through a magnetic undulator~\cite{kondratenko1980generating,huang2007review}. This operation mode is used as the fundamental operating mode at various FEL facilities worldwide~\cite{ackermann2007operation,emma2010first,ishikawa2012compact,kang2017hard,zhao2017sxfel,decking2020mhz,wang2023physical}. It has become an indispensable tool for cutting-edge experimental applications across diverse research areas, such as fundamental physics~\cite{kayser2019core}, material science~\cite{forte2024resonant}, structural biology~\cite{chapman2011femtosecond} and ultra-fast chemistry~\cite{kupper2014x}.

In SASE FEL, the amplification process is initiated by shot noise in the electron beam, which is fundamentally stochastic. As a result, the radiation emitted by SASE FEL also exhibits stochastic properties. The effective design and planning of user equipment and experiments are heavily reliant on a thorough understanding of the properties of the radiation pulses produced. Consequently, comprehensive investigations have been conducted for the statistical properties of SASE FEL, including pulse energy, temporal and spectral distribution and pulse duration ~\cite{saldin1998statistical,yurkov2002statistical,bermudez2021study}. Among the properties of interest, pulse energy is a fundamental parameter that is essential for the proper development of experimental methods in FEL ~\cite{bernstein2009near}. In particular, when conducting intensity dependent measurements, it is necessary that pulse energy is measured in parallel with the experimental data acquisition process and in a non-intrusive manner. Therefore, to enhance reproducibility and reduce the need for extensive post-processing~\cite{tanaka2009high}, efforts have been made to minimize pulse energy fluctuations~\cite{shintake2009stable}.

This paper presents the investigation of a robust setup to improve the pulse energy stability of the SASE process. The method involves introducing a dispersive section to control the fluctuation of the bunching factor and thereby improving the stability of the radiation pulse energy generated in downstream undulators. The dispersive section can be a magnetic chicane, an off-resonant undulator, or a similar component that can introduce longitudinal dispersion. A chicane is chosen in this study due to its flexibility in tuning dispersive strength. As will be demonstrated, compact chicanes that can be integrated into the break sections between undulators provide sufficient dispersive strength for the intended application. 

Analysis of SASE radiation shows that in the exponential growth regime both longitudinal and transverse coherence properties are improved, reaching their optimum levels at the onset of the saturation regime and declines thereafter~\cite{bermudez2021study,saldin2008coherence}. However, pulse energy fluctuations increase during the exponential growth phase, peaking just before saturation, and then decrease in the post-saturation regime.  In systems designed to avoid saturation to maintain coherence properties, pulse energy fluctuations can become substantial when the lasing process is terminated before saturation. Additionally, when FEL facilities operate at their shortest achievable wavelengths, the length of the undulator line may be insufficient for full saturation. Therefore, to address such scenarios, the present study examines the impact of inserting a chicane during the exponential growth phase of the SASE process. At this stage, the energy modulation along the electron bunch is sufficiently developed to enable effective manipulation of the bunching distributions both within each individual electron bunch and across multiple bunches. This manipulation directly influences the lasing process in downstream undulators, resulting in less dispersed pulse energy distributions~\cite{thompson2018possible}. Furthermore, in advanced schemes employing a two-stage lasing process, where the first stage operates in SASE mode, careful control of the radiation intensity at this stage may be required. This control serves two purposes: first, to preserve the key properties of the radiation in the initial stage, and second, to prevent excessive energy spread from degrading the lasing process in the subsequent stage. Given these needs, the first-stage lasing can be terminated in the exponential growth regime to maintain moderate levels of bunching and energy spread~\cite{hara2013two,Sun, guo2024experimental}. In such advanced schemes, the implementation of a chicane is crucial for time delay and bunching control, and it can also be beneficial for stabilizing pulse energy fluctuations in the second stage.

Finally, it is worth noting that the insertion of a dispersive section, such as a magnetic chicane, between undulators is usually referred to as optical klystron (OK) configuration~\cite{bonifacio1992theory}. Most analyses of the OK setup applied to SASE focus on increasing the rate of intensity gain~\cite{ding2006optical,geloni2021revision}. In practical applications, significant intensity enhancement can be achieved for longer wavelengths, such as those in the UV regime~\cite{penco2015experimental}. For shorter wavelengths in the X-ray region, multiple chicanes must be employed and moderate growth rate enhancements and reductions in saturation length are achieved ~\cite{penco2017optical,prat2021demonstration,kittel2024enhanced}. In both single and multiple chicane configurations, the dispersive section is usually positioned starting from the early exponential regime to accelerate the bunching process. In contrast, the present objective of enhancing stability requires the dispersive section to be situated deeper into the exponential regime. This allows for more effective utilization of the established energy modulation, and thereby, enabling better control and containment of the bunching distribution. 


The paper is organized as follows. Section II presents the stabilization scheme for the SASE process. The stabilization effect is demonstrated based on three-dimensional SASE simulations for X-ray SASE FELs. Section III analyzes the general stability effect of the chicane using a simplified model, followed by a detailed investigation of its impact on radiation properties through three-dimensional SASE simulations. Section IV demonstrates the chicane’s stabilization effect in a two-stage lasing process designed to produce two-color single-spike pulses. Finally, section V provides the concluding remarks.

\section{The stabilization scheme}

In the lasing process of FEL, high intensity radiation is generated by coherently radiating electrons. The radiation wavelength ($\lambda_{r}$) is determined by the electron bunch and undulator parameters and it satisfies the resonance condition:
\begin{equation}
\lambda_{r}=\frac{\lambda_{u}}{2\gamma^{2}}(1+\frac{K^{2}}{2}),
\end{equation}
where $\gamma$ is the electron energy in units of electron rest mass, while $\lambda_{u}$ and $K$ are the undulator period and undulator parameter, respectively. During the lasing process, electrons periodically form microbunches within slices of length $\lambda_{r}$. This periodic density modulation is quantified by the bunching factor $b=|\Sigma e^{i\theta}|/N$ , where $\theta$ represents the electron phase (ranging from 0 to 2$\pi$) and $N$ is the total number of electrons in each slice. Since radiation intensity is directly linked to the degree of bunching, controlling the bunching factor distribution allows for modulation of the emitted radiation energy in subsequent undulators.

For SASE FEL, the radiation process initiates from a white noise electron density spectrum. As the electron bunches propagate through the undulator, the initial random bunching factor within each bunch is progressively amplified. 
In the exponential gain regime, the bunching factor fluctuations tend to be large, leading to radiation pulses with power distributions characterized by random spikes in both position and intensity. This results in significant shot-to-shot variations in pulse energy~\cite{bermudez2021study}.

\begin{figure*}[hbt!]
    \centering
    \includegraphics[width=\textwidth]{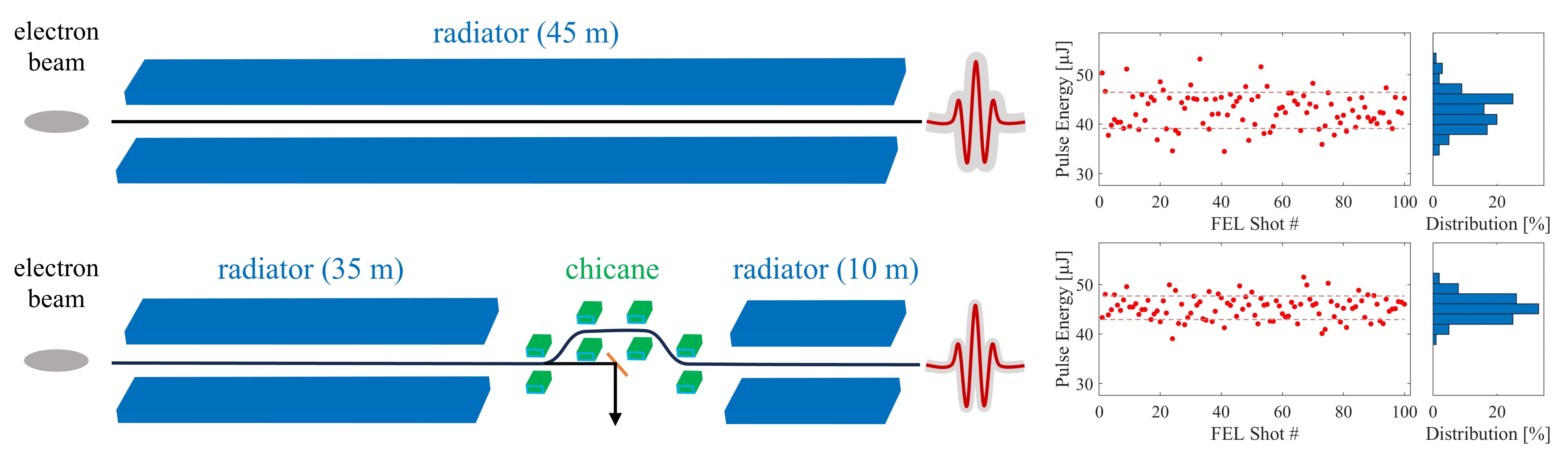}
    \caption{Illustration of using a chicane for achieving stabilization effect of the pulse energy in a SASE FEL.}
    \label{fig1}
\end{figure*}

The fluctuations can be mitigated through a dispersive section, and the stabilization effect that can be attained through the use of a chicane is demonstrated in Fig.~\ref{fig1}, with SASE simulations performed with the GENESIS 1.3 code~\cite{reiche1999genesis}. These simulations are based on the machine and beam parameters of the proposed Shenzhen Superconducting Soft X-ray Free Electron Laser (S\textsuperscript{3}FEL) ~\cite{wang2023physical}, which are detailed in Table~\ref{table:1}. Ideal electron bunches with Gaussian current profiles are assumed, and radiation at a wavelength of $\lambda_r$=1 nm is considered. A total of 100 numerical simulations are performed, each with a different random seed. The chicane has a total length of 0.5 m, allowing it to be placed in the 1-meter-long break section between undulator modules. The compact chicane provides sufficient dispersion strength for the purpose of controlling the bunching distributions~\cite{prat2016undulator}. For the simulation results presented in Fig.~\ref{fig1}, the chosen dispersive strength is $R_{56}=0.8~{\upmu}m$, corresponding to a delay of approximately 1.3 fs between the electron bunch and the radiation pulse. The chicane is positioned in the break section following the seventh undulator module (near the midpoint of the exponential regime), and pulse energy distributions are evaluated after the ninth undulator module (toward the end of the exponential regime).
 \begin{table}[h!]
\centering
\caption{SASE simulation parameters.}
\begin{tabular}[t]{lc} 
 \hline\hline
 Parameters & Value\\ 
 \hline
 Beam energy [GeV] & 2.5  \\ 
 Peak curren t [A]& 800  \\
 $\epsilon_{x,n}/\epsilon_{y,n}$ [mm$\cdot$mrad] & 0.4/0.4 \\
 $\sigma_E$ [keV] & 180 \\
 Undulator period length [cm] & 3.0 \\
 Undulator period number & 133 \\[1ex] 
 \hline\hline
\end{tabular}
\label{table:1}
\end{table}

\hfill

Significant reduction in the shot-to-shot fluctuation of the pulse energy ($E_{r}$) can be achieved with the use of the chicane. Compared to the normal SASE process, after travelling the same distance of 45 m along the undulator line, the relative fluctuation (defined as the ratio of the standard deviation to the mean and referred to as the fluctuation in the following text) of the pulse energy decreased from $\sigma_{E_{r}}/\overline{E_{r}}=8.55\%$ to $\sigma_{E_{r}}/\overline{E_{r}}=5.25\%$. The average pulse energy is less influenced by the chicane and increased from 42.77 ${\upmu}$J to 45.32 ${\upmu}$J.

In this study, radiation emitted from upstream undulators is effectively blocked to prevent interference with the lasing process in the downstream undulators. This setup allows for the analysis of modified bunching distributions and their resultant impact on the radiation characteristics in the downstream undulators.


\section{Principle of the method }

In this section, the stabilization effect of the chicane is first illustrated through a simplified model that utilizes a energy modulation to density modulation transformation equation. Then, the impact on radiation characteristics is evaluated with SASE simulations.

\subsection{Stability effect of the chicane with a simplified model}


In the SASE FEL process, energy and density modulations gradually develop from shot noise as the electron bunch propagates through the undulator. Over each resonant wavelength, the energy modulation acquires a sinusoidal shape, with its amplitude represented by the energy offset $\Delta E$, which varies randomly along the bunch. Despite these fluctuations, a Gaussian current profile results in energy modulation profiles that are Gaussian-shaped, with fluctuating peak amplitudes and widths. To convert the Gaussian-shaped energy modulation profiles into density modulation via a chicane, the formalism used in the high-gain harmonic generation (HGHG) process is adopted, where a dispersive section performs a similar function \cite{yu1991generation}. In HGHG, an external laser imprints a periodic energy modulation on an electron bunch within a modulator undulator, and a subsequent dispersive section transforms this modulation into a density modulation, following: 
\begin{equation}
b_{h}=exp(-\frac{1}{2}h^{2}B^{2})J_{h}(hAB),
\end{equation}
where $A=\Delta E/\sigma_{E}$ is the relative energy modulation amplitude and $B=(2\pi/\lambda)(\sigma_{E}/E)R_{56}$ is related to the longitudinal dispersion. $R_{56}$ is the momentum compaction of the chicane and $h$ is the harmonic number. In this simplified model, the case of $h=1$ is considered and the pulse energy distribution is evaluated according to the distribution of the integral $\int b^2 dt$.

\begin{figure}[htb!]
    \centering
    \includegraphics[width=1\columnwidth]{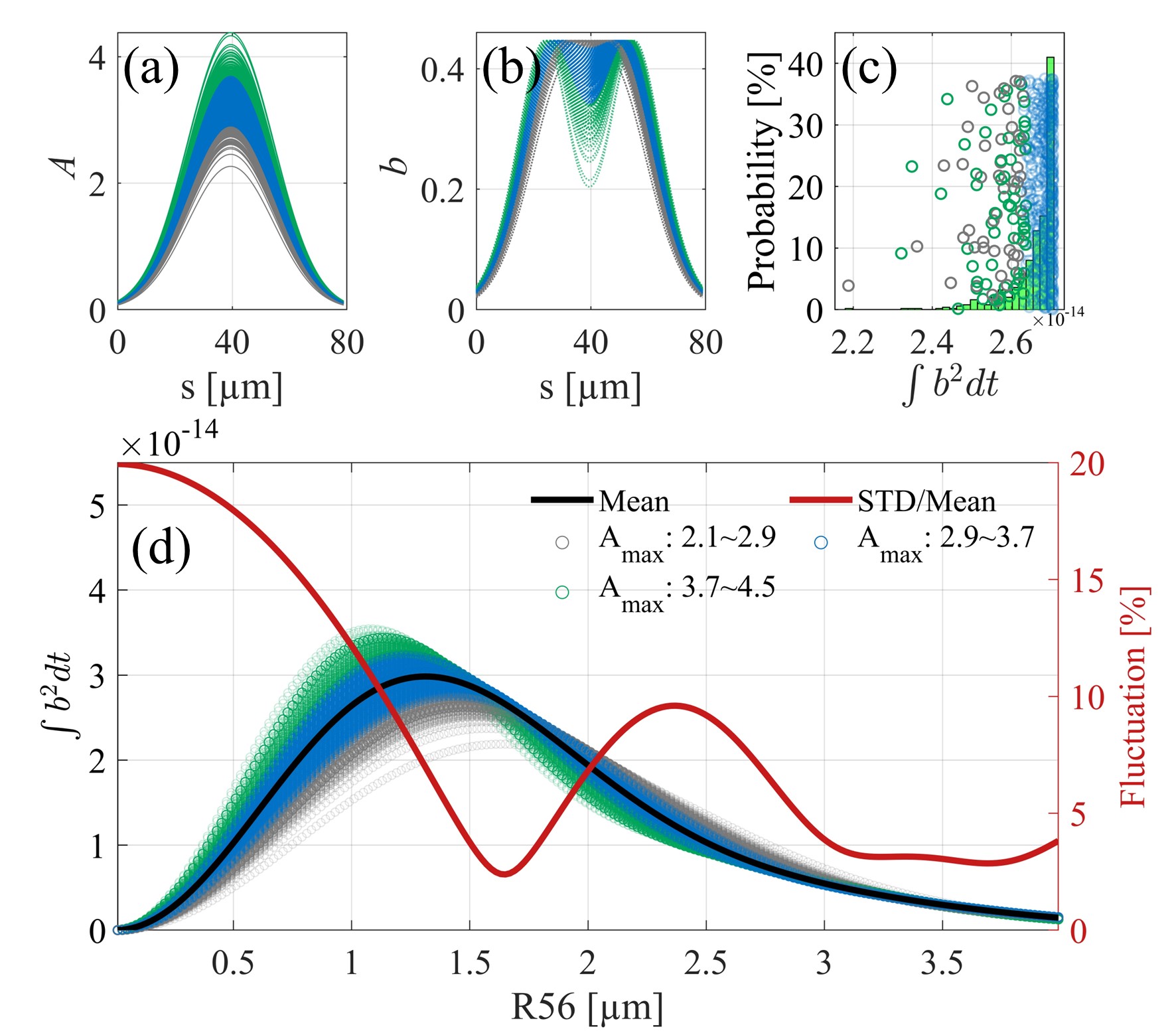}
    \caption{The effect of the chicane on randomly generated Gaussian energy modulation amplitudes: (a) energy modulation amplitudes with Gaussian profiles that vary in their maximum values; (b) the resulting bunching profiles; (c) distribution of $\int b^2 dt$ for a chicane strength of $R_{56}=1.65~{\upmu}m$; and (d) the distributions of $\int b^2 dt$ for different $R_{56}$ values of the chicane.}
    \label{fig2}
\end{figure}

 The randomly generated energy modulation amplitude profiles are shown in Fig.~\ref{fig2}(a). The maximum values of the Gaussian curves, $A_{max}$, are distributed according to a gamma distribution with a mean of 3.3 and a standard deviation of 0.33. The width of the Gaussian curves is fixed. The resulting bunching factor $b$ is calculated for a wavelength of $\lambda_r=1$ nm and with the relative energy spread $\sigma_{E}/E=7\times10^{-5}$.

First, the bunching manipulation due to the chicane is examined for a fixed dispersive strength where $R_{56}=1.66~{\upmu}m$. The resulting bunching profiles are shown in Fig.~\ref{fig2}(b), and the distribution of $\int b^2 dt$ is shown in Fig.~\ref{fig2}(c). For shots with low levels of $A$ (grey curves), the bunching factor profiles maintain a Gaussian shape, with slightly lower $b^2$ integrals (grey dots). As $A$ increases (from blue to green curves), the bunching factors exhibit increasingly damped behavior at the central peak. This suppression of the bunching distribution limits the further increase of $b^2$ integrals, causing them to plateau around a maximum value (blue dots) and eventually decrease as $A$ values become large (green dots). The overall effect is that the bunching profiles and the $\int b^2 dt$ distributions are contained, resulting in a more concentrated pulse energy distribution. The relative fluctuation of the $b^2$ integrals, quantified by $\sigma_{\int b_{i}^2 dt}/<\int b_{i}^2 dt>_{i}$ ($\sigma$ represents the standard deviation of the integrals, and $<\dotsm>_i$ denotes the average of the different shots $i$), is 2.3\%.

To compare the fluctuation with a normal SASE process, the distribution of the $\int A^2 dt$ is used. The case with $R_{56}=0~{\upmu}m$ corresponds to the normal SASE process. In this simple model, however, inserting $R_{56}=0$ into Eq. (2) leads to zero bunching profiles, making it unsuitable for evaluating the fluctuation. Despite this, in the exponential regime of the SASE process, the bunching profile along the electron bunch is strongly correlated with the profile of $A$. Therefore, before the chicane, the bunching profiles can be approximated by scaling the $A$ profiles with a scaling factor. As a result, the relative fluctuation in pulse energy ($\sigma_{\int b_{i}^2 dt}/<\int b_{i}^2 dt>_{i}$) is expected to be similar to $\sigma_{\int A_{i}^2 dt}/<\int A_{i}^2 dt>_{i}$. For the Gaussian $A$ profiles shown in Fig.~\ref{fig2}(a), the relative fluctuation of $\int A^2 dt$ is ~20\%. This suggests that the introduction of a chicane improves the stability of pulse energy fluctuations compared to the normal SASE case without the chicane.

To further investigate how the stability effect varies with different values of $R_{56}$, a scan of $R_{56}$ is performed to assess the dependence of pulse energy fluctuation on the dispersion strength. The different distributions of the resulting bunching factor integrals (or equivalently, the pulse energy distributions) are shown in Fig.~\ref{fig2}(d). The relative fluctuation starts near 20\% (close to the fluctuation level discussed for the normal SASE process) for $R_{56}$ values just above 0. As the dispersive strength increases, the fluctuation initially decreases but begins to rise once the dispersion becomes excessively strong. Nevertheless, across all tested $R_{56}$ values, the relative fluctuation remains below 20\%, thereby demonstrating the stabilizing enhancement effect of the chicane. The average pulse energy shows an initial increase and followed by monotonic decrease. Notably, the highest average pulse energy and the smallest relative fluctuation of the energy are obtained at different $R_{56}$ values. For $R_{56}=1.34~\upmu m$, the highest average pulse energy is reached, and the relative fluctuation in this case is $6.5\%$. Increasing $R_{56}$ to $1.66~\upmu m$, the average pulse energy decreases, but the minimum relative fluctuation of $2.3\%$ may be achieved.

\subsection{Simulation results}


The effect of the chicane on the SASE process is analyzed through three-dimensional simulations using the same system setup and beam parameters as presented in section II.

Before entering the chicane, the energy modulation amplitude and the bunching factor fluctuate randomly along the electron bunch. Across different shots, the fluctuations in the integral $\int A^2 dt$ closely resembles those in $\int b^2 dt$, with respective values of approximately 14.2\% and 13.2\%. The close agreement between these values demonstrates the strong correlation between $A$ and $b$, as noted in previous discussions. Despite the fluctuations along the bunch, the overall shapes of the energy modulations follow that of the (Gaussian) current distributions. The fitted Gaussian curves for $A$ exhibit variations in their maximum values, along with slight changes in the width of the curves. Given these distributions, the general trend of the chicane’s effect with increasing $R_{56}$, as described by the simple model above, is expected to be reproduced. However, the exact values of the relative fluctuations are not anticipated to match precisely, as the Gaussian $A$ profiles used in the previous section were generated rather arbitrarily.

\begin{figure}[htb!]
    \centering
    \includegraphics[width=0.9\linewidth]{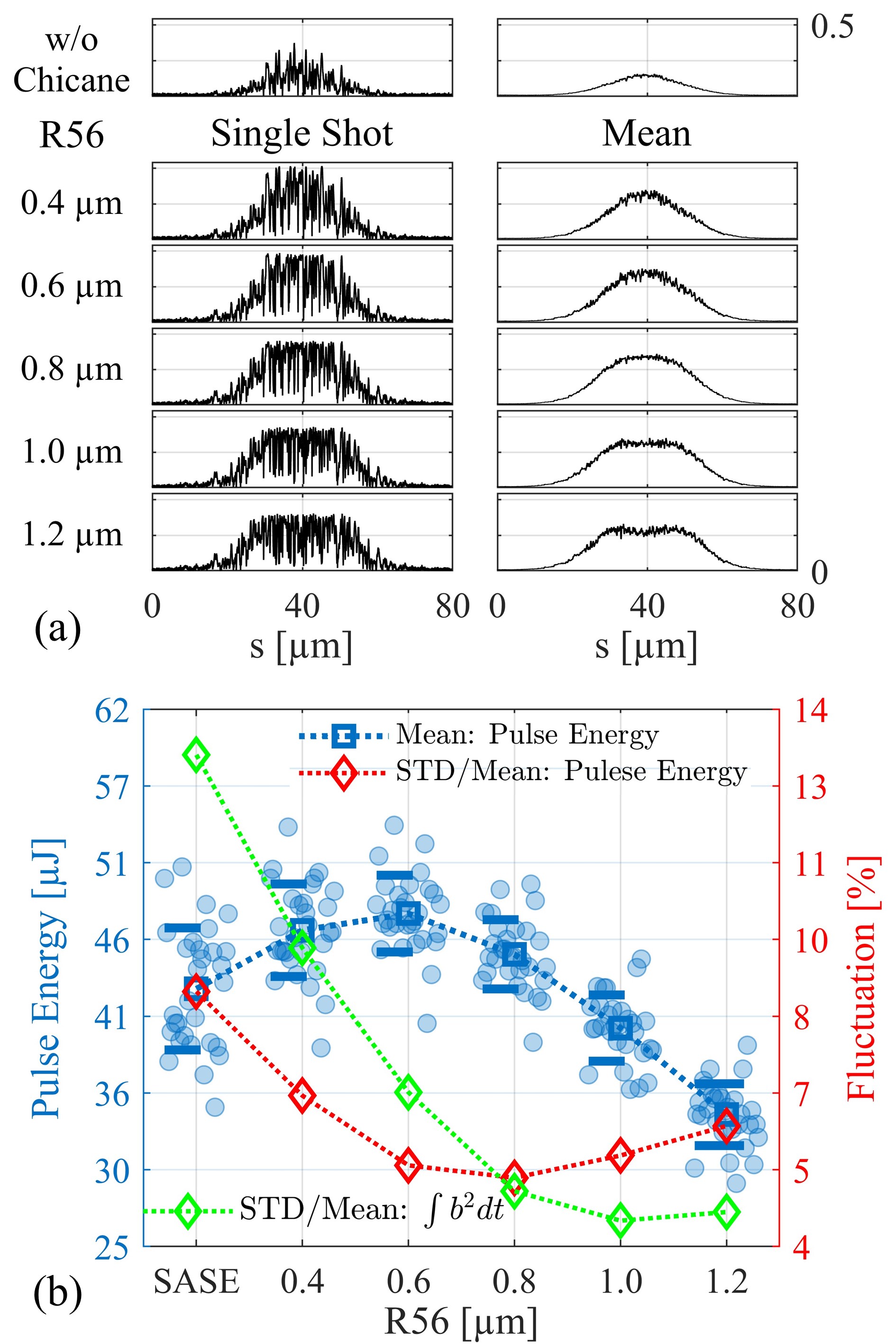}
    \caption{Effect of the chicane with increasing $R_{56}$ value: (a) the bunching factor distribution along the electron bunch for individual shots (left panels) and the mean of the shots (right panels); and (b) the distributions of the pulse energy and their relative fluctuations.}
    \label{fig3}
\end{figure}

Figure ~\ref{fig3}(a) shows the effect of chicane strength on the bunching factor distributions for an individual shot (left panels) and the mean of all shots (right panels). At low chicane strength, such as $R_{56}=0.4~{\upmu}m$, the bunching factor increases compared to the normal SASE case. As $R_{56}$ increases, the maximum achievable bunching factor reaches its peak, after which it decreases in regions with high $A$ values due to the overstretching of the electron longitudinal distributions within each slice. This containment of the bunching factor is reflected in the flattened peaks of the bunching profiles. For the mean bunching distributions, the central regions of $b$ initially plateau and then experience a slight suppression as $R_{56}$ increases. The fluctuations of the bunching integrals with increasing $R_{56}$ are shown in Fig.~\ref{fig3}(b) (green curve), where a general trend of an initial decrease followed by a slight increase is observed.

The fluctuation of the pulse energy radiated after the ninth undulator, as shown by the red line in Fig.~\ref{fig3}(b), is influenced by the modulated bunching distributions. The trend for the fluctuation of the pulse energy, as predicted by the simple model, is similarly observed here. The most significant stabilization effect occurs at $R_{56}=0.8~{\upmu}m$, where the relative fluctuation is 5.3\%. The mean pulse energy follows the bell-shaped pattern predicted in Fig.~\ref{fig2}(d), initially increasing with smaller $R_{56}$ values, peaking at $R_{56}=0.6~\upmu m$, and then decreasing as the central region becomes over-bunched. The progression from the first maximum average to the most stabilized point is consistently reproduced. The discrepancy between the fluctuations in pulse energy and the bunching integral arises from the influence of other factors affecting the lasing process, such as energy spread and additional three-dimensional effects of the electron beam and the radiated pulse.

It is also noted that, compared to the simple model, a lower $R_{56}$ value is required for optimal stabilization effect in the SASE process. This is because a moderate level of bunching has developed in conjunction with the formation of energy modulation, and thus requires less dispersion to contain the bunching distributions.

\begin{figure}[htb!]
    \centering
    \includegraphics[width=1.0\linewidth]{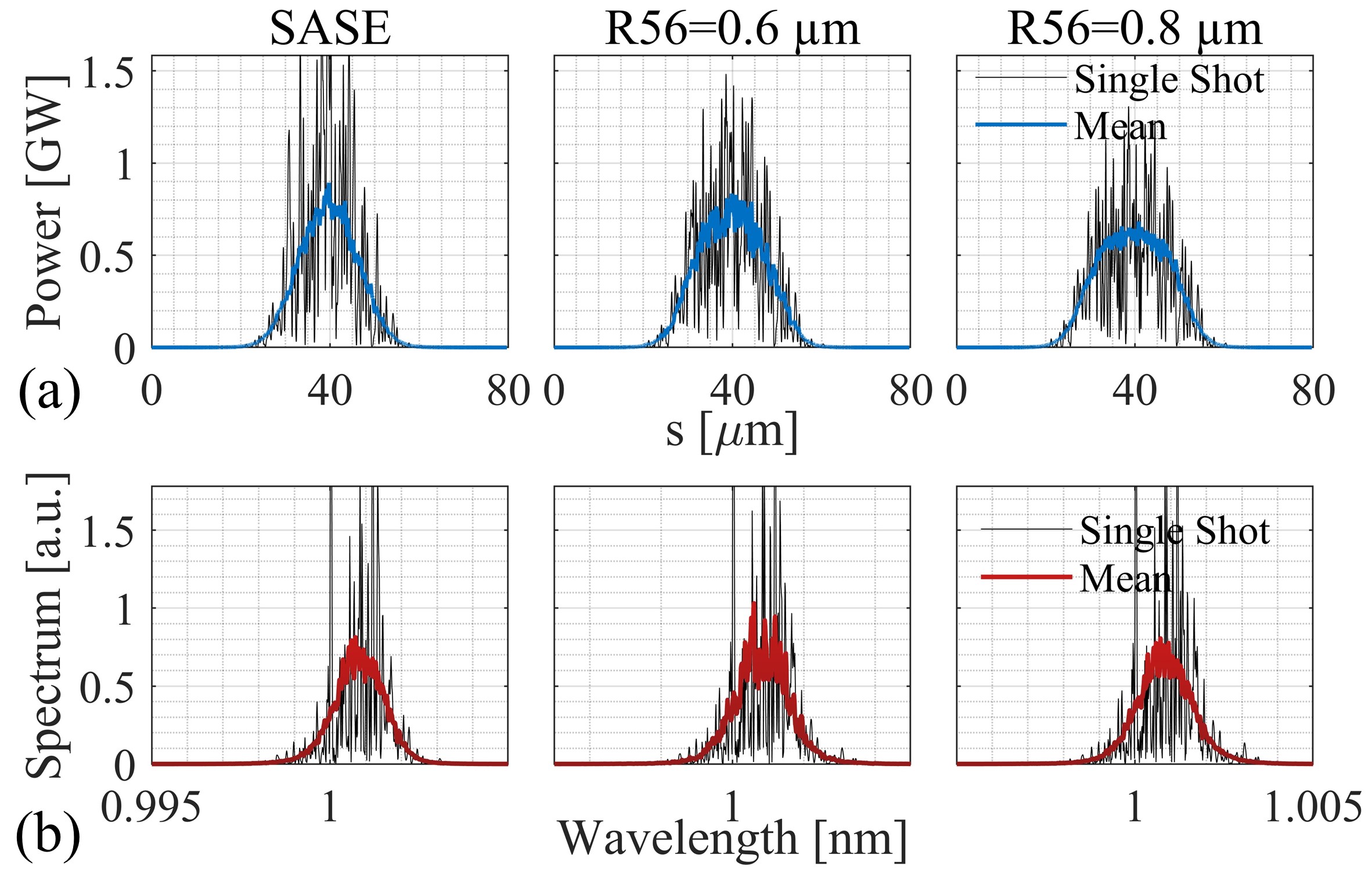}
    \caption{The (a) temporal and (b) spectral profiles of the radiation pulses for the normal SASE case and the chicane inserted cases.}
    \label{fig4}
\end{figure}

Figure~\ref{fig4} shows the power profile and spectrum of the radiated pulse at the end of the undulator line for both the normal SASE case and the stabilized cases, with $R_{56}=0.6~\upmu m$ and $R_{56}=0.8~\upmu m$. Due to the effect of the chicane, the bunching near the tails of the Gaussian current profile increases, causing a larger portion of the electron bunch to radiate and resulting in an increased pulse duration.

The spectrum bandwidth increases slightly from 0.160\% for the case without the chicane to 0.172\% and 0.174\% for cases with the chicane included. This is due to the growth of longitudinal modes within the radiation pulse. As shown in Fig.~\ref{fig4}(a), the average number of spikes in the power profiles rises from 58 in the normal SASE case to 70 and 78 for the cases with chicanes at $R_{56}=0.6~\upmu m$ and $R_{56}=0.8~\upmu m$, respectively.

Combining the influence of the chicane on the modulation of the bunching factor within each shot and across the shots, it is expected that radiation pulses with single-spike characteristics will exhibit enhanced stability without the negative effects of bandwidth broadening. This is demonstrated in the following section through an example that uses two-stage lasing processes, where the first stage operates in the SASE mode.

\section{Stabilization Application}

In two-color generation schemes that employ a two-stage lasing process, the chicane can control the time delay between the pulses generated in each stage. In addition, the chicane plays a key role in manipulating the bunching of the electron beam, enabling the generation of high-intensity pulses in the second stage within a short undulator length. In these schemes, if the first stage operates in the SASE mode, the chicane may also introduce a stabilizing effect on the pulse generated in the second stage.

The stabilizing effect has been demonstrated in the two-color scheme proposed by Sun et al.~\cite{Sun}. This scheme makes use of a dual-chirped electron beam (positive and negative energy chirp generated through interaction with a few-cycle laser in a wiggler) and two tapered undulator sections to produce ultra-short two-color pulses. The photon energies of the two pulses are $\omega$ and $\omega+\Delta\omega$, where $\Delta\omega/{\omega} \thicksim 4\%$. The chicane is placed between the two sections to separate the pulses by a few femtoseconds. Notably, the pulse length and peak power in the second section exhibit significantly more stable behavior compared to those from the first undulator section, due to the influence of the chicane.

Another scheme, developed at LCLS ~\cite{guo2024experimental}, also utilizes a two-stage lasing process to generate short pulses at the fundamental frequency $\omega$ and its second harmonic $2\omega$ in the first and second undulator sections, respectively. This process relies on an electron bunch with a high current spike—resulting from the shaping of the photocathode laser pulse and amplified through the acceleration and compression process, together with a positive energy chirp created by a magnetic wiggler. The time delay control between the two pulses is accomplished by inserting a chicane between the two undulator sections. Here, the stabilization effect of the chicane is further examined with numerical simulations performed with GENESIS 1.3.

\begin{figure}[htbp]
    \centering
    \includegraphics[width=1\linewidth]{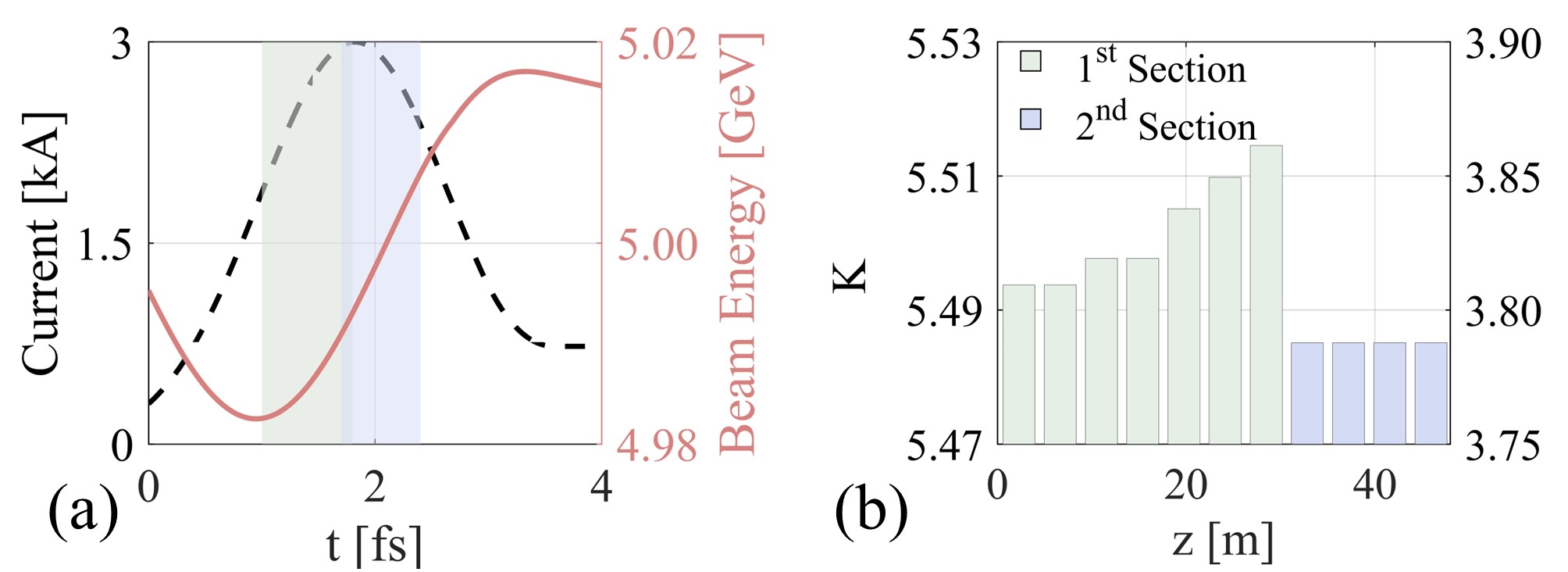}
    \caption{Parameters used in the simulations for the two-color scheme: (a) the current and energy profile of the ideal electron bunch where the shaded areas show the lasing sections along the bunch for the first and second pulses and (b) the undulator $K$ values for both sections.}
    \label{fig5}
\end{figure}

The simulations are based on ideal electron bunches that possess the current and energy chirp profiles described in~\cite{guo2024experimental}. The longitudinal profiles for the current and energy are shown in Fig.~\ref{fig5}(a). The other uniform bunch parameters and undulator parameters used in the numerical simulations are detailed in Table~\ref{table:2}. 

 \begin{table}[h!]
\centering
\caption{Simulation parameters for the two-color scheme.}
\begin{tabular}[t]{lc} 
 \hline\hline
 Parameters & Value\\ 
 \hline
 Beam energy [GeV] & 5.0  \\ 
 Peak current [kA]& 3  \\
 $\epsilon_{x,n}/\epsilon_{y,n}$ [mm$\cdot$mrad] & 1.17/0.45 \\
 $\sigma_E$ [keV] & 500 \\
 Undulator period length [cm] & 3.9 \\
 Undulator period number & 89 \\[1ex] 
 \hline\hline
\end{tabular}
\label{table:2}
\end{table}

 

The taper configuration in the first undulator section is optimized to maintain a moderate bunching factor as the electron bunch exits this section. An example of such a configuration is shown in Fig.~\ref{fig5}(b). In this setup, the taper of the last three undulators in the first section is adjusted to match the energy chirp gradient along the left shoulder of the current peak (Fig.~\ref{fig5}(a), green area), generating the first ($\omega$ frequency) pulse at a photon energy around 370 eV. The undulators in the second section are set to constant $K$ values, producing second color pulse ($2\omega$ frequency) with approximately 740 eV photon energy, by utilizing the right shoulder of the current peak (Fig.~\ref{fig5}(a), blue area). A total of 100 simulations are performed by varying the random seed for each. A small chicane, with a total length of 0.5 m, is selected as it provides sufficient dispersive strength and time delay, and is inserted into the 1-meter-long break section between the two undulator sections.
 
\begin{figure}[htb!]
    \centering
    \includegraphics[width=1\linewidth]{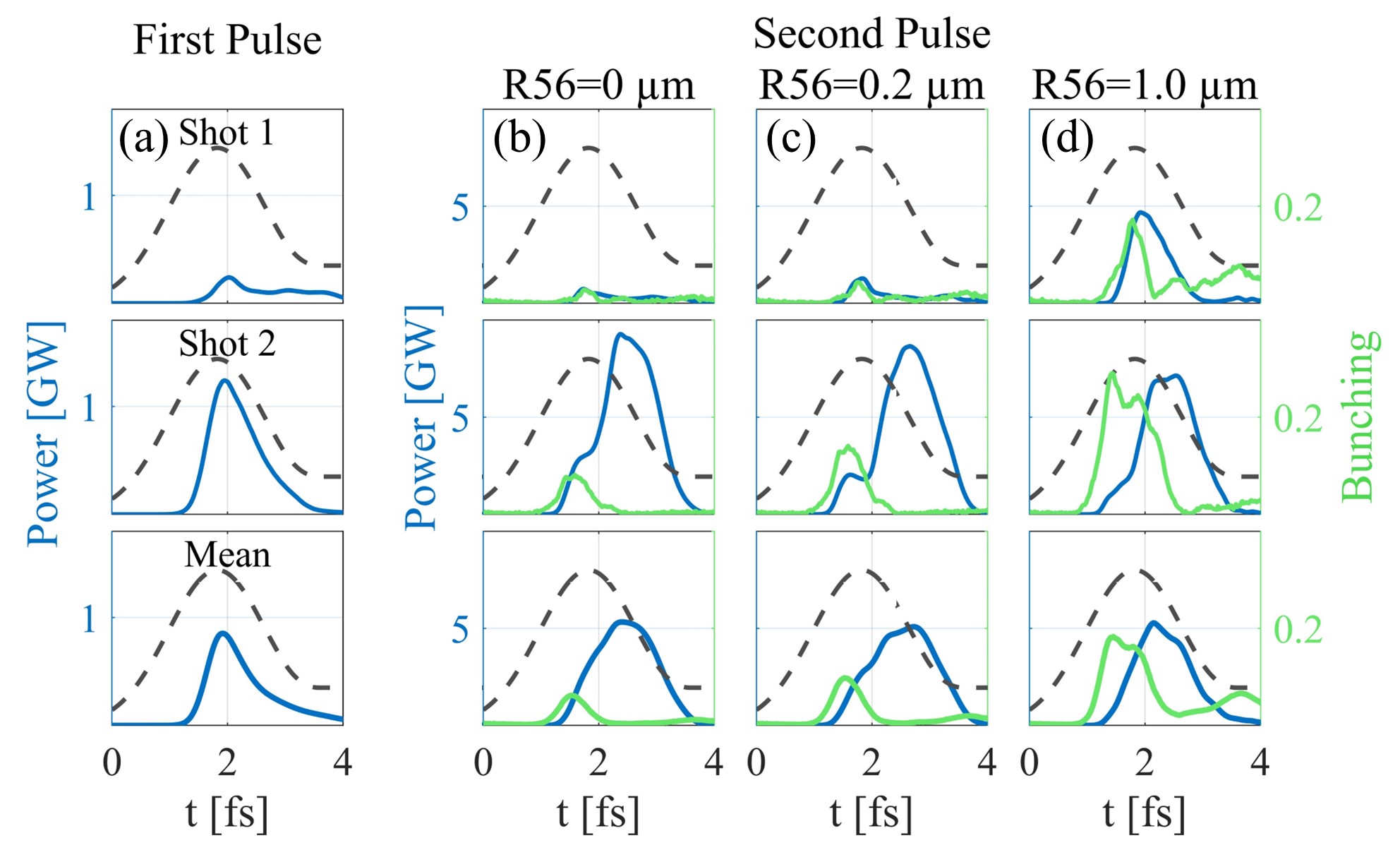}
    \caption{Temporal distribution of the pulses generated in the (a) first undulator section shown for two individual shots (top and middle rows) and the average of all shots (bottom row). Bunching factor distributions after traversing the chicane and the resulting second ($2\omega$) pulses generated with chicane strengths of: (b) $R_{56} = 0~{\upmu}m$, (c) $R_{56}=0.2~{\upmu}m$ and (d) $R_{56}=1.0~{\upmu}m$. The black dashed curve represents the current.}
    \label{fig6}
\end{figure}

The power profiles of the first pulses generated in the first undulator section are shown in Fig.~\ref{fig6}(a), for two individual shots (top and middle rows) and the average of all shots (bottom row). The pulses exhibit attosecond-duration single-spike characteristics. Fig.~\ref{fig6}(b) presents the bunching at the second harmonic (green lines) after the first undulator section, in the absence of the chicane. The two single shot cases shown in Fig.~\ref{fig6}(a) and Fig.~\ref{fig6}(b) illustrate that, due to the SASE process, the peak power of the first pulses fluctuates significantly, and large fluctuations are also observed in the bunching factor. Consequently, without the chicane, the second ($2\omega$) pulses generated in the second undulator section also exhibit large fluctuations (shown in blue curves of Fig.~\ref{fig6}(b)). However, the introduction of the chicane modulates the bunching levels, and leads to more consistent power generation in the second undulator section, as shown in Fig.~\ref{fig6}(c) and Fig.~\ref{fig6}(d) for a small chicane strengths of $R_{56}=0.2~\upmu m$ (0.33 fs delay) and a larger strength at $R_{56}=1.0~\upmu m$ (1.67 fs delay).

Compared to the case without the chicane, the primary effect of a weak chicane strength is an increase in the bunching factor for shots with initially low bunching levels, resulting in a slight increase in power (compare Fig.~\ref{fig6}(b) and Fig.~\ref{fig6}(c), top row). Shots with high bunching levels are less affected by the weak chicane, as seen in the middle row of Fig.~\ref{fig6}(c). 

As the chicane strength increases, fluctuations in both the bunching factor and the corresponding power distributions of the second pulses are significantly reduced (Fig.~\ref{fig6}(d)). Notably, shots with low bunching factors show a substantial increase in both the bunching factor and the power of the second pulse. Meanwhile, shots with high bunching levels from the first undulator section are capped at around 0.3, suppressing power generation in the second section. This balance between low and high power shots leads to a more uniform power distribution of the second pulses.

The impact on the average bunching profiles is shown in the bottom row of Fig.~\ref{fig6}. For a small value of $R_{56}$, the average bunching is raised. At higher $R_{56}$ values, however, this increase is suppressed, which is consistent with that observed in standard SASE simulations discussed in the previous section.

\begin{figure}[htb!]
    \centering
    \includegraphics[width=1\linewidth]{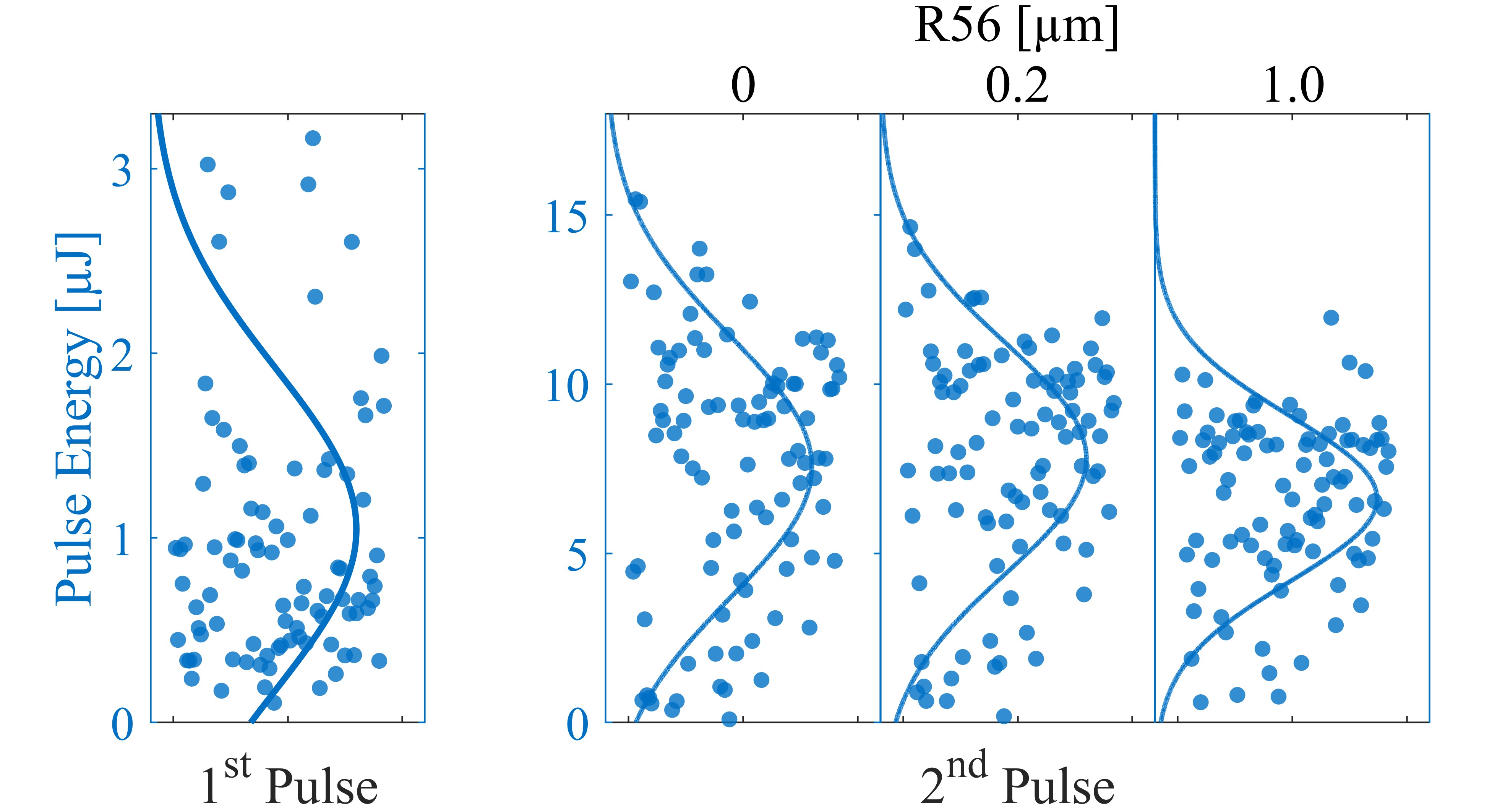}
    \caption{The pulse energy distributions of the first ($\omega$) pulses and the second ($2\omega$) pulses for chicane strengths of $R_{56} = 0 ~{\upmu}m$, $R_{56}=0.2~{\upmu}m$, and $R_{56}=1.0~{\upmu}m$}
    \label{fig7}
\end{figure}

The pulse energy distributions for the first pulses and the second pulses at different chicane strengths are shown in Fig.~\ref{fig7}. The data points represent the pulse energy of individual shots, while the curves are Gaussian fits to the energy distributions. For the first pulses, significant fluctuations are evident from the widely spread energy points and the large width of the fitted curve. For this pulse, the average energy is 0.93 $\upmu$J, with a standard deviation of 0.75 $\upmu$J, resulting in a relative fluctuation of 80.3\%.

For the second pulses, energy fluctuations are notably reduced with the use of a chicane. Without the chicane, the average pulse energy is 7.26 $\upmu$J, with a relative fluctuation of 63.0\%. After implementing the chicane, the average energy increases slightly to 7.36 $\upmu$J at $R_{56}=0.2 ~{\upmu}m$, primarily due to the increase in energy for lower-energy shots. The relative fluctuation decreases to 55.2\%, indicating an improvement in pulse energy stability compared to the case without the chicane, even at a minimal chicane strength. At a chicane strength of $R_{56}=1.0 ~{\upmu}m$, due to the suppression of high bunching profiles, the average energy slightly decreases to 6.29 $\upmu$J while the relative fluctuation further reduces to 43.1\%. For the current undulator setup, this relative fluctuation value is the optimal stability achieved by scanning chicane $R_{56}$ from $0.2~{\upmu}m$ to $1.2~{\upmu}m$ in steps of $0.2~{\upmu}m$.

\begin{figure}[htbp]
    \centering
    \includegraphics[width=1\linewidth]{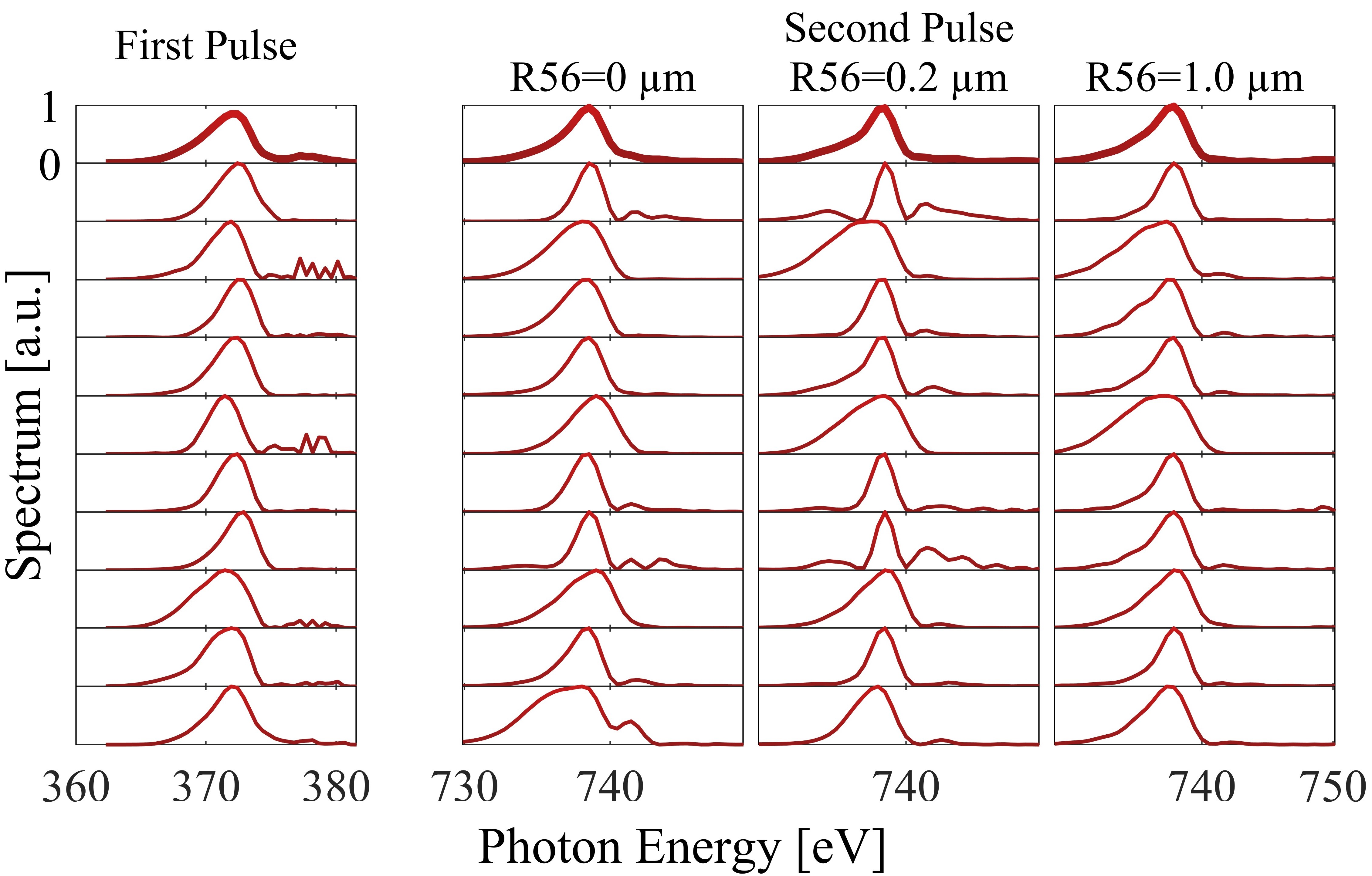}
    \caption{The spectral distribution of the first ($\omega$) pulses and the second ($2\omega$) pulses with different chicane strengths.Top row: the mean of the spectrum distributions; lower rows: individual shots.}
    \label{fig8}
\end{figure}

The spectral properties of the first and second pulses are shown in Fig.~\ref{fig8}. The spectrum of the first pulse exhibits a single peak around 370 eV, with a mean spectral FWHM of 3.67 eV. For the second pulses, unlike the typical SASE case, the introduction of the chicane does not significantly alter the bandwidth. Without the chicane, the average FWHM is 3.46 eV. With the chicane inserted, the average FWHM is 3.04 eV for $R_{56}=0.2~{\upmu}m$ and 3.48 eV for $R_{56}=1.0~{\upmu}m$.

\section{Summary}

This paper presents a robust methodology for reducing pulse energy fluctuations in FEL lasing schemes that involve the SASE process. The approach employs a small chicane during the exponential growth phase of the SASE process to manipulate the bunching distributions, accumulated in upstream undulators, both along and across the electron bunches. The effect of the chicane on pulse energy distribution is investigated through numerical simulations of a normal SASE process. Different $R_{56}$ values of the chicane are tested and the results show that the chicane effectively reduces pulse energy fluctuations. The trend in stability with increasing chicane strength aligns with predictions from a simple theoretical model. However, a slight increase in bandwidth is noted due to the formation of additional power spikes in the long SASE pulses. The stability effect is further demonstrated in two-stage undulator setups for generating two-color pulses, where the first stage operates in SASE mode. In this context, the chicane is shown to enhance pulse energy stability without significantly affecting the bandwidth. The study shows that the simple chicane setup can be an effective method for optimizing pulse energy stability in SASE-based lasing schemes.

\begin{acknowledgments}
This work is supported by the Scientific Instrument Developing Project of Chinese Academy of Sciences (Grant No. GJJSTD20220001) and the National Natural Science Foundation of China (Grant No. 22288201).
\end{acknowledgments}


\bibliography{main}

\providecommand{\noopsort}[1]{}\providecommand{\singleletter}[1]{#1}%
\begin{thebibliography}{34}%
\makeatletter
\providecommand \@ifxundefined [1]{%
 \@ifx{#1\undefined}
}%
\providecommand \@ifnum [1]{%
 \ifnum #1\expandafter \@firstoftwo
 \else \expandafter \@secondoftwo
 \fi
}%
\providecommand \@ifx [1]{%
 \ifx #1\expandafter \@firstoftwo
 \else \expandafter \@secondoftwo
 \fi
}%
\providecommand \natexlab [1]{#1}%
\providecommand \enquote  [1]{``#1''}%
\providecommand \bibnamefont  [1]{#1}%
\providecommand \bibfnamefont [1]{#1}%
\providecommand \citenamefont [1]{#1}%
\providecommand \href@noop [0]{\@secondoftwo}%
\providecommand \href [0]{\begingroup \@sanitize@url \@href}%
\providecommand \@href[1]{\@@startlink{#1}\@@href}%
\providecommand \@@href[1]{\endgroup#1\@@endlink}%
\providecommand \@sanitize@url [0]{\catcode `\\12\catcode `\$12\catcode `\&12\catcode `\#12\catcode `\^12\catcode `\_12\catcode `\%12\relax}%
\providecommand \@@startlink[1]{}%
\providecommand \@@endlink[0]{}%
\providecommand \url  [0]{\begingroup\@sanitize@url \@url }%
\providecommand \@url [1]{\endgroup\@href {#1}{\urlprefix }}%
\providecommand \urlprefix  [0]{URL }%
\providecommand \Eprint [0]{\href }%
\providecommand \doibase [0]{http://dx.doi.org/}%
\providecommand \selectlanguage [0]{\@gobble}%
\providecommand \bibinfo  [0]{\@secondoftwo}%
\providecommand \bibfield  [0]{\@secondoftwo}%
\providecommand \translation [1]{[#1]}%
\providecommand \BibitemOpen [0]{}%
\providecommand \bibitemStop [0]{}%
\providecommand \bibitemNoStop [0]{.\EOS\space}%
\providecommand \EOS [0]{\spacefactor3000\relax}%
\providecommand \BibitemShut  [1]{\csname bibitem#1\endcsname}%
\let\auto@bib@innerbib\@empty
\bibitem [{\citenamefont {Kondratenko}\ and\ \citenamefont {Saldin}(1980)}]{kondratenko1980generating}%
  \BibitemOpen
  \bibfield  {author} {\bibinfo {author} {\bibfnamefont {AM}~\bibnamefont {Kondratenko}}\ and\ \bibinfo {author} {\bibfnamefont {EL}~\bibnamefont {Saldin}},\ }\bibfield  {title} {\enquote {\bibinfo {title} {Generating of coherent radiation by a relativistic electron beam in an ondulator},}\ }\href@noop {} {\bibfield  {journal} {\bibinfo  {journal} {Part. Accel.}\ }\textbf {\bibinfo {volume} {10}},\ \bibinfo {pages} {207--216} (\bibinfo {year} {1980})}\BibitemShut {NoStop}%
\bibitem [{\citenamefont {Huang}\ and\ \citenamefont {Kim}(2007)}]{huang2007review}%
  \BibitemOpen
  \bibfield  {author} {\bibinfo {author} {\bibfnamefont {Zhirong}\ \bibnamefont {Huang}}\ and\ \bibinfo {author} {\bibfnamefont {Kwang-Je}\ \bibnamefont {Kim}},\ }\bibfield  {title} {\enquote {\bibinfo {title} {Review of x-ray free-electron laser theory},}\ }\href@noop {} {\bibfield  {journal} {\bibinfo  {journal} {Phys. Rev. ST Accel. Beams}\ }\textbf {\bibinfo {volume} {10}},\ \bibinfo {pages} {034801} (\bibinfo {year} {2007})}\BibitemShut {NoStop}%
\bibitem [{\citenamefont {Ackermann}\ \emph {et~al.}(2007)\citenamefont {Ackermann}, \citenamefont {Asova}, \citenamefont {Ayvazyan}, \citenamefont {Azima}, \citenamefont {Baboi}, \citenamefont {B{\"a}hr}, \citenamefont {Balandin}, \citenamefont {Beutner}, \citenamefont {Brandt}, \citenamefont {Bolzmann} \emph {et~al.}}]{ackermann2007operation}%
  \BibitemOpen
  \bibfield  {author} {\bibinfo {author} {\bibfnamefont {W~al}\ \bibnamefont {Ackermann}}, \bibinfo {author} {\bibfnamefont {G}~\bibnamefont {Asova}}, \bibinfo {author} {\bibfnamefont {V}~\bibnamefont {Ayvazyan}}, \bibinfo {author} {\bibfnamefont {A}~\bibnamefont {Azima}}, \bibinfo {author} {\bibfnamefont {Nicoleta}\ \bibnamefont {Baboi}}, \bibinfo {author} {\bibfnamefont {J}~\bibnamefont {B{\"a}hr}}, \bibinfo {author} {\bibfnamefont {Vladimir}\ \bibnamefont {Balandin}}, \bibinfo {author} {\bibfnamefont {Bolko}\ \bibnamefont {Beutner}}, \bibinfo {author} {\bibfnamefont {A}~\bibnamefont {Brandt}}, \bibinfo {author} {\bibfnamefont {Andy}\ \bibnamefont {Bolzmann}},  \emph {et~al.},\ }\bibfield  {title} {\enquote {\bibinfo {title} {Operation of a free-electron laser from the extreme ultraviolet to the water window},}\ }\href@noop {} {\bibfield  {journal} {\bibinfo  {journal} {Nat. Photonics}\ }\textbf {\bibinfo {volume} {1}},\ \bibinfo {pages} {336--342} (\bibinfo {year} {2007})}\BibitemShut {NoStop}%
\bibitem [{\citenamefont {Emma}\ \emph {et~al.}(2010)\citenamefont {Emma}, \citenamefont {Akre}, \citenamefont {Arthur}, \citenamefont {Bionta}, \citenamefont {Bostedt}, \citenamefont {Bozek}, \citenamefont {Brachmann}, \citenamefont {Bucksbaum}, \citenamefont {Coffee}, \citenamefont {Decker} \emph {et~al.}}]{emma2010first}%
  \BibitemOpen
  \bibfield  {author} {\bibinfo {author} {\bibfnamefont {Paul}\ \bibnamefont {Emma}}, \bibinfo {author} {\bibfnamefont {R}~\bibnamefont {Akre}}, \bibinfo {author} {\bibfnamefont {J}~\bibnamefont {Arthur}}, \bibinfo {author} {\bibfnamefont {R}~\bibnamefont {Bionta}}, \bibinfo {author} {\bibfnamefont {C}~\bibnamefont {Bostedt}}, \bibinfo {author} {\bibfnamefont {J}~\bibnamefont {Bozek}}, \bibinfo {author} {\bibfnamefont {A}~\bibnamefont {Brachmann}}, \bibinfo {author} {\bibfnamefont {P}~\bibnamefont {Bucksbaum}}, \bibinfo {author} {\bibfnamefont {Ryan}\ \bibnamefont {Coffee}}, \bibinfo {author} {\bibfnamefont {F-J}\ \bibnamefont {Decker}},  \emph {et~al.},\ }\bibfield  {title} {\enquote {\bibinfo {title} {First lasing and operation of an {\aa}ngstrom-wavelength free-electron laser},}\ }\href@noop {} {\bibfield  {journal} {\bibinfo  {journal} {Nat. Photonics}\ }\textbf {\bibinfo {volume} {4}},\ \bibinfo {pages} {641--647} (\bibinfo {year} {2010})}\BibitemShut {NoStop}%
\bibitem [{\citenamefont {Ishikawa}\ \emph {et~al.}(2012)\citenamefont {Ishikawa}, \citenamefont {Aoyagi}, \citenamefont {Asaka}, \citenamefont {Asano}, \citenamefont {Azumi}, \citenamefont {Bizen}, \citenamefont {Ego}, \citenamefont {Fukami}, \citenamefont {Fukui}, \citenamefont {Furukawa} \emph {et~al.}}]{ishikawa2012compact}%
  \BibitemOpen
  \bibfield  {author} {\bibinfo {author} {\bibfnamefont {Tetsuya}\ \bibnamefont {Ishikawa}}, \bibinfo {author} {\bibfnamefont {Hideki}\ \bibnamefont {Aoyagi}}, \bibinfo {author} {\bibfnamefont {Takao}\ \bibnamefont {Asaka}}, \bibinfo {author} {\bibfnamefont {Yoshihiro}\ \bibnamefont {Asano}}, \bibinfo {author} {\bibfnamefont {Noriyoshi}\ \bibnamefont {Azumi}}, \bibinfo {author} {\bibfnamefont {Teruhiko}\ \bibnamefont {Bizen}}, \bibinfo {author} {\bibfnamefont {Hiroyasu}\ \bibnamefont {Ego}}, \bibinfo {author} {\bibfnamefont {Kenji}\ \bibnamefont {Fukami}}, \bibinfo {author} {\bibfnamefont {Toru}\ \bibnamefont {Fukui}}, \bibinfo {author} {\bibfnamefont {Yukito}\ \bibnamefont {Furukawa}},  \emph {et~al.},\ }\bibfield  {title} {\enquote {\bibinfo {title} {A compact x-ray free-electron laser emitting in the sub-{\aa}ngstr{\"o}m region},}\ }\href@noop {} {\bibfield  {journal} {\bibinfo  {journal} {Nat. Photonics}\ }\textbf {\bibinfo {volume} {6}},\ \bibinfo {pages} {540--544} (\bibinfo {year} {2012})}\BibitemShut
  {NoStop}%
\bibitem [{\citenamefont {Kang}\ \emph {et~al.}(2017)\citenamefont {Kang}, \citenamefont {Min}, \citenamefont {Heo}, \citenamefont {Kim}, \citenamefont {Yang}, \citenamefont {Kim}, \citenamefont {Nam}, \citenamefont {Baek}, \citenamefont {Choi}, \citenamefont {Mun} \emph {et~al.}}]{kang2017hard}%
  \BibitemOpen
  \bibfield  {author} {\bibinfo {author} {\bibfnamefont {Heung-Sik}\ \bibnamefont {Kang}}, \bibinfo {author} {\bibfnamefont {Chang-Ki}\ \bibnamefont {Min}}, \bibinfo {author} {\bibfnamefont {Hoon}\ \bibnamefont {Heo}}, \bibinfo {author} {\bibfnamefont {Changbum}\ \bibnamefont {Kim}}, \bibinfo {author} {\bibfnamefont {Haeryong}\ \bibnamefont {Yang}}, \bibinfo {author} {\bibfnamefont {Gyujin}\ \bibnamefont {Kim}}, \bibinfo {author} {\bibfnamefont {Inhyuk}\ \bibnamefont {Nam}}, \bibinfo {author} {\bibfnamefont {Soung~Youl}\ \bibnamefont {Baek}}, \bibinfo {author} {\bibfnamefont {Hyo-Jin}\ \bibnamefont {Choi}}, \bibinfo {author} {\bibfnamefont {Geonyeong}\ \bibnamefont {Mun}},  \emph {et~al.},\ }\bibfield  {title} {\enquote {\bibinfo {title} {Hard x-ray free-electron laser with femtosecond-scale timing jitter},}\ }\href@noop {} {\bibfield  {journal} {\bibinfo  {journal} {Nat. Photonics}\ }\textbf {\bibinfo {volume} {11}},\ \bibinfo {pages} {708--713} (\bibinfo {year} {2017})}\BibitemShut {NoStop}%
\bibitem [{\citenamefont {Zhao}\ \emph {et~al.}(2017)\citenamefont {Zhao}, \citenamefont {Wang}, \citenamefont {Gu}, \citenamefont {Yin}, \citenamefont {Fang}, \citenamefont {Gu}, \citenamefont {Leng}, \citenamefont {Zhou}, \citenamefont {Liu}, \citenamefont {Tang} \emph {et~al.}}]{zhao2017sxfel}%
  \BibitemOpen
  \bibfield  {author} {\bibinfo {author} {\bibfnamefont {Zhentang}\ \bibnamefont {Zhao}}, \bibinfo {author} {\bibfnamefont {Dong}\ \bibnamefont {Wang}}, \bibinfo {author} {\bibfnamefont {Qiang}\ \bibnamefont {Gu}}, \bibinfo {author} {\bibfnamefont {Lixin}\ \bibnamefont {Yin}}, \bibinfo {author} {\bibfnamefont {Guoping}\ \bibnamefont {Fang}}, \bibinfo {author} {\bibfnamefont {Ming}\ \bibnamefont {Gu}}, \bibinfo {author} {\bibfnamefont {Yongbin}\ \bibnamefont {Leng}}, \bibinfo {author} {\bibfnamefont {Qiaogen}\ \bibnamefont {Zhou}}, \bibinfo {author} {\bibfnamefont {Bo}~\bibnamefont {Liu}}, \bibinfo {author} {\bibfnamefont {Chuanxiang}\ \bibnamefont {Tang}},  \emph {et~al.},\ }\bibfield  {title} {\enquote {\bibinfo {title} {Sxfel: a soft x-ray free electron laser in china},}\ }\href@noop {} {\bibfield  {journal} {\bibinfo  {journal} {Synchrotron Radiat. News}\ }\textbf {\bibinfo {volume} {30}},\ \bibinfo {pages} {29--33} (\bibinfo {year} {2017})}\BibitemShut {NoStop}%
\bibitem [{\citenamefont {Decking}\ \emph {et~al.}(2020)\citenamefont {Decking}, \citenamefont {Abeghyan}, \citenamefont {Abramian}, \citenamefont {Abramsky}, \citenamefont {Aguirre}, \citenamefont {Albrecht}, \citenamefont {Alou}, \citenamefont {Altarelli}, \citenamefont {Altmann}, \citenamefont {Amyan} \emph {et~al.}}]{decking2020mhz}%
  \BibitemOpen
  \bibfield  {author} {\bibinfo {author} {\bibfnamefont {Winfried}\ \bibnamefont {Decking}}, \bibinfo {author} {\bibfnamefont {S}~\bibnamefont {Abeghyan}}, \bibinfo {author} {\bibfnamefont {P}~\bibnamefont {Abramian}}, \bibinfo {author} {\bibfnamefont {A}~\bibnamefont {Abramsky}}, \bibinfo {author} {\bibfnamefont {A}~\bibnamefont {Aguirre}}, \bibinfo {author} {\bibfnamefont {C}~\bibnamefont {Albrecht}}, \bibinfo {author} {\bibfnamefont {P}~\bibnamefont {Alou}}, \bibinfo {author} {\bibfnamefont {M}~\bibnamefont {Altarelli}}, \bibinfo {author} {\bibfnamefont {P}~\bibnamefont {Altmann}}, \bibinfo {author} {\bibfnamefont {K}~\bibnamefont {Amyan}},  \emph {et~al.},\ }\bibfield  {title} {\enquote {\bibinfo {title} {A mhz-repetition-rate hard x-ray free-electron laser driven by a superconducting linear accelerator},}\ }\href@noop {} {\bibfield  {journal} {\bibinfo  {journal} {Nat. Photonics}\ }\textbf {\bibinfo {volume} {14}},\ \bibinfo {pages} {391--397} (\bibinfo {year} {2020})}\BibitemShut {NoStop}%
\bibitem [{\citenamefont {Wang}\ \emph {et~al.}(2023)\citenamefont {Wang}, \citenamefont {Zeng}, \citenamefont {Shao}, \citenamefont {Liang}, \citenamefont {Yi}, \citenamefont {Yu}, \citenamefont {Sun}, \citenamefont {Li}, \citenamefont {Feng}, \citenamefont {Wang} \emph {et~al.}}]{wang2023physical}%
  \BibitemOpen
  \bibfield  {author} {\bibinfo {author} {\bibfnamefont {Xiaofan}\ \bibnamefont {Wang}}, \bibinfo {author} {\bibfnamefont {Li}~\bibnamefont {Zeng}}, \bibinfo {author} {\bibfnamefont {Jiahang}\ \bibnamefont {Shao}}, \bibinfo {author} {\bibfnamefont {Yifan}\ \bibnamefont {Liang}}, \bibinfo {author} {\bibfnamefont {Huaiqian}\ \bibnamefont {Yi}}, \bibinfo {author} {\bibfnamefont {Yong}\ \bibnamefont {Yu}}, \bibinfo {author} {\bibfnamefont {Jitao}\ \bibnamefont {Sun}}, \bibinfo {author} {\bibfnamefont {Xinmeng}\ \bibnamefont {Li}}, \bibinfo {author} {\bibfnamefont {Chao}\ \bibnamefont {Feng}}, \bibinfo {author} {\bibfnamefont {Zhen}\ \bibnamefont {Wang}},  \emph {et~al.},\ }\bibfield  {title} {\enquote {\bibinfo {title} {Physical design for shenzhen superconducting soft x-ray free-electron laser (s3fel)},}\ }\href@noop {} {\bibfield  {journal} {\bibinfo  {journal} {in Proceedings of the International Particle Accelerator Conference (IPAC’23), Venice, Italy}\ ,\ \bibinfo {pages} {TUPL043}} (\bibinfo {year}
  {2023})}\BibitemShut {NoStop}%
\bibitem [{\citenamefont {Kayser}\ \emph {et~al.}(2019)\citenamefont {Kayser}, \citenamefont {Milne}, \citenamefont {Jurani{\'c}}, \citenamefont {Sala}, \citenamefont {Czapla-Masztafiak}, \citenamefont {Follath}, \citenamefont {Kav{\v{c}}i{\v{c}}}, \citenamefont {Knopp}, \citenamefont {Rehanek}, \citenamefont {B{\l}achucki} \emph {et~al.}}]{kayser2019core}%
  \BibitemOpen
  \bibfield  {author} {\bibinfo {author} {\bibfnamefont {Yves}\ \bibnamefont {Kayser}}, \bibinfo {author} {\bibfnamefont {Chris}\ \bibnamefont {Milne}}, \bibinfo {author} {\bibfnamefont {Pavle}\ \bibnamefont {Jurani{\'c}}}, \bibinfo {author} {\bibfnamefont {Leonardo}\ \bibnamefont {Sala}}, \bibinfo {author} {\bibfnamefont {Joanna}\ \bibnamefont {Czapla-Masztafiak}}, \bibinfo {author} {\bibfnamefont {Rolf}\ \bibnamefont {Follath}}, \bibinfo {author} {\bibfnamefont {Matja{\v{z}}}\ \bibnamefont {Kav{\v{c}}i{\v{c}}}}, \bibinfo {author} {\bibfnamefont {Gregor}\ \bibnamefont {Knopp}}, \bibinfo {author} {\bibfnamefont {Jens}\ \bibnamefont {Rehanek}}, \bibinfo {author} {\bibfnamefont {Wojciech}\ \bibnamefont {B{\l}achucki}},  \emph {et~al.},\ }\bibfield  {title} {\enquote {\bibinfo {title} {Core-level nonlinear spectroscopy triggered by stochastic x-ray pulses},}\ }\href@noop {} {\bibfield  {journal} {\bibinfo  {journal} {Nat. Commun}\ }\textbf {\bibinfo {volume} {10}},\ \bibinfo {pages} {4761} (\bibinfo {year}
  {2019})}\BibitemShut {NoStop}%
\bibitem [{\citenamefont {Forte}\ \emph {et~al.}(2024)\citenamefont {Forte}, \citenamefont {Gawne}, \citenamefont {Alaa El-Din}, \citenamefont {Humphries}, \citenamefont {Preston}, \citenamefont {Cr{\'e}pisson}, \citenamefont {Campbell}, \citenamefont {Svensson}, \citenamefont {Azadi}, \citenamefont {Heighway} \emph {et~al.}}]{forte2024resonant}%
  \BibitemOpen
  \bibfield  {author} {\bibinfo {author} {\bibfnamefont {Alessandro}\ \bibnamefont {Forte}}, \bibinfo {author} {\bibfnamefont {Thomas}\ \bibnamefont {Gawne}}, \bibinfo {author} {\bibfnamefont {Karim~K}\ \bibnamefont {Alaa El-Din}}, \bibinfo {author} {\bibfnamefont {Oliver~S}\ \bibnamefont {Humphries}}, \bibinfo {author} {\bibfnamefont {Thomas~R}\ \bibnamefont {Preston}}, \bibinfo {author} {\bibfnamefont {C{\'e}line}\ \bibnamefont {Cr{\'e}pisson}}, \bibinfo {author} {\bibfnamefont {Thomas}\ \bibnamefont {Campbell}}, \bibinfo {author} {\bibfnamefont {Pontus}\ \bibnamefont {Svensson}}, \bibinfo {author} {\bibfnamefont {Sam}\ \bibnamefont {Azadi}}, \bibinfo {author} {\bibfnamefont {Patrick}\ \bibnamefont {Heighway}},  \emph {et~al.},\ }\bibfield  {title} {\enquote {\bibinfo {title} {Resonant inelastic x-ray scattering in warm-dense fe compounds beyond the sase fel resolution limit},}\ }\href@noop {} {\bibfield  {journal} {\bibinfo  {journal} {Commun. Phys}\ }\textbf {\bibinfo {volume} {7}},\ \bibinfo {pages}
  {266} (\bibinfo {year} {2024})}\BibitemShut {NoStop}%
\bibitem [{\citenamefont {Chapman}\ \emph {et~al.}(2011)\citenamefont {Chapman}, \citenamefont {Fromme}, \citenamefont {Barty}, \citenamefont {White}, \citenamefont {Kirian}, \citenamefont {Aquila}, \citenamefont {Hunter}, \citenamefont {Schulz}, \citenamefont {DePonte}, \citenamefont {Weierstall} \emph {et~al.}}]{chapman2011femtosecond}%
  \BibitemOpen
  \bibfield  {author} {\bibinfo {author} {\bibfnamefont {Henry~N}\ \bibnamefont {Chapman}}, \bibinfo {author} {\bibfnamefont {Petra}\ \bibnamefont {Fromme}}, \bibinfo {author} {\bibfnamefont {Anton}\ \bibnamefont {Barty}}, \bibinfo {author} {\bibfnamefont {Thomas~A}\ \bibnamefont {White}}, \bibinfo {author} {\bibfnamefont {Richard~A}\ \bibnamefont {Kirian}}, \bibinfo {author} {\bibfnamefont {Andrew}\ \bibnamefont {Aquila}}, \bibinfo {author} {\bibfnamefont {Mark~S}\ \bibnamefont {Hunter}}, \bibinfo {author} {\bibfnamefont {Joachim}\ \bibnamefont {Schulz}}, \bibinfo {author} {\bibfnamefont {Daniel~P}\ \bibnamefont {DePonte}}, \bibinfo {author} {\bibfnamefont {Uwe}\ \bibnamefont {Weierstall}},  \emph {et~al.},\ }\bibfield  {title} {\enquote {\bibinfo {title} {Femtosecond x-ray protein nanocrystallography},}\ }\href@noop {} {\bibfield  {journal} {\bibinfo  {journal} {Nature}\ }\textbf {\bibinfo {volume} {470}},\ \bibinfo {pages} {73--77} (\bibinfo {year} {2011})}\BibitemShut {NoStop}%
\bibitem [{\citenamefont {K{\"u}pper}\ \emph {et~al.}(2014)\citenamefont {K{\"u}pper}, \citenamefont {Stern}, \citenamefont {Holmegaard}, \citenamefont {Filsinger}, \citenamefont {Rouz{\'e}e}, \citenamefont {Rudenko}, \citenamefont {Johnsson}, \citenamefont {Martin}, \citenamefont {Adolph}, \citenamefont {Aquila} \emph {et~al.}}]{kupper2014x}%
  \BibitemOpen
  \bibfield  {author} {\bibinfo {author} {\bibfnamefont {Jochen}\ \bibnamefont {K{\"u}pper}}, \bibinfo {author} {\bibfnamefont {Stephan}\ \bibnamefont {Stern}}, \bibinfo {author} {\bibfnamefont {Lotte}\ \bibnamefont {Holmegaard}}, \bibinfo {author} {\bibfnamefont {Frank}\ \bibnamefont {Filsinger}}, \bibinfo {author} {\bibfnamefont {Arnaud}\ \bibnamefont {Rouz{\'e}e}}, \bibinfo {author} {\bibfnamefont {Artem}\ \bibnamefont {Rudenko}}, \bibinfo {author} {\bibfnamefont {Per}\ \bibnamefont {Johnsson}}, \bibinfo {author} {\bibfnamefont {Andrew~V}\ \bibnamefont {Martin}}, \bibinfo {author} {\bibfnamefont {Marcus}\ \bibnamefont {Adolph}}, \bibinfo {author} {\bibfnamefont {Andrew}\ \bibnamefont {Aquila}},  \emph {et~al.},\ }\bibfield  {title} {\enquote {\bibinfo {title} {X-ray diffraction from isolated and strongly aligned gas-phase molecules with a free-electron laser},}\ }\href@noop {} {\bibfield  {journal} {\bibinfo  {journal} {Phys. Rev. Lett.}\ }\textbf {\bibinfo {volume} {112}},\ \bibinfo {pages} {083002}
  (\bibinfo {year} {2014})}\BibitemShut {NoStop}%
\bibitem [{\citenamefont {Saldin}\ \emph {et~al.}(1998)\citenamefont {Saldin}, \citenamefont {Schneidmiller},\ and\ \citenamefont {Yurkov}}]{saldin1998statistical}%
  \BibitemOpen
  \bibfield  {author} {\bibinfo {author} {\bibfnamefont {Evgenij~L}\ \bibnamefont {Saldin}}, \bibinfo {author} {\bibfnamefont {Evgeny~A}\ \bibnamefont {Schneidmiller}}, \ and\ \bibinfo {author} {\bibfnamefont {MV}~\bibnamefont {Yurkov}},\ }\bibfield  {title} {\enquote {\bibinfo {title} {Statistical properties of radiation from vuv and x-ray free electron laser},}\ }\href@noop {} {\bibfield  {journal} {\bibinfo  {journal} {Opt. Commun.}\ }\textbf {\bibinfo {volume} {148}},\ \bibinfo {pages} {383--403} (\bibinfo {year} {1998})}\BibitemShut {NoStop}%
\bibitem [{\citenamefont {Yurkov}(2002)}]{yurkov2002statistical}%
  \BibitemOpen
  \bibfield  {author} {\bibinfo {author} {\bibfnamefont {MV}~\bibnamefont {Yurkov}},\ }\bibfield  {title} {\enquote {\bibinfo {title} {Statistical properties of sase fel radiation: experimental results from the vuv fel at the tesla test facility at desy},}\ }\href@noop {} {\bibfield  {journal} {\bibinfo  {journal} {Nucl. Instrum. Methods Phys. Res., Sect. A}\ }\textbf {\bibinfo {volume} {483}},\ \bibinfo {pages} {51--56} (\bibinfo {year} {2002})}\BibitemShut {NoStop}%
\bibitem [{\citenamefont {Berm{\'u}dez~Macias}\ \emph {et~al.}(2021)\citenamefont {Berm{\'u}dez~Macias}, \citenamefont {D{\"u}sterer}, \citenamefont {Ivanov}, \citenamefont {Liu}, \citenamefont {Brenner}, \citenamefont {R{\"o}nsch-Schulenburg}, \citenamefont {Czwalinna},\ and\ \citenamefont {Yurkov}}]{bermudez2021study}%
  \BibitemOpen
  \bibfield  {author} {\bibinfo {author} {\bibfnamefont {Ivette~J}\ \bibnamefont {Berm{\'u}dez~Macias}}, \bibinfo {author} {\bibfnamefont {Stefan}\ \bibnamefont {D{\"u}sterer}}, \bibinfo {author} {\bibfnamefont {Rosen}\ \bibnamefont {Ivanov}}, \bibinfo {author} {\bibfnamefont {Jia}\ \bibnamefont {Liu}}, \bibinfo {author} {\bibfnamefont {G{\"u}nter}\ \bibnamefont {Brenner}}, \bibinfo {author} {\bibfnamefont {Juliane}\ \bibnamefont {R{\"o}nsch-Schulenburg}}, \bibinfo {author} {\bibfnamefont {Marie~K}\ \bibnamefont {Czwalinna}}, \ and\ \bibinfo {author} {\bibfnamefont {Mikhail~V}\ \bibnamefont {Yurkov}},\ }\bibfield  {title} {\enquote {\bibinfo {title} {Study of temporal, spectral, arrival time and energy fluctuations of sase fel pulses},}\ }\href@noop {} {\bibfield  {journal} {\bibinfo  {journal} {Opt. Express}\ }\textbf {\bibinfo {volume} {29}},\ \bibinfo {pages} {10491--10508} (\bibinfo {year} {2021})}\BibitemShut {NoStop}%
\bibitem [{\citenamefont {Bernstein}\ \emph {et~al.}(2009)\citenamefont {Bernstein}, \citenamefont {Acremann}, \citenamefont {Scherz}, \citenamefont {Burkhardt}, \citenamefont {St{\"o}hr}, \citenamefont {Beye}, \citenamefont {Schlotter}, \citenamefont {Beeck}, \citenamefont {Sorgenfrei}, \citenamefont {Pietzsch} \emph {et~al.}}]{bernstein2009near}%
  \BibitemOpen
  \bibfield  {author} {\bibinfo {author} {\bibfnamefont {DP}~\bibnamefont {Bernstein}}, \bibinfo {author} {\bibfnamefont {Y}~\bibnamefont {Acremann}}, \bibinfo {author} {\bibfnamefont {A}~\bibnamefont {Scherz}}, \bibinfo {author} {\bibfnamefont {M}~\bibnamefont {Burkhardt}}, \bibinfo {author} {\bibfnamefont {J}~\bibnamefont {St{\"o}hr}}, \bibinfo {author} {\bibfnamefont {M}~\bibnamefont {Beye}}, \bibinfo {author} {\bibfnamefont {WF}~\bibnamefont {Schlotter}}, \bibinfo {author} {\bibfnamefont {T}~\bibnamefont {Beeck}}, \bibinfo {author} {\bibfnamefont {F}~\bibnamefont {Sorgenfrei}}, \bibinfo {author} {\bibfnamefont {A}~\bibnamefont {Pietzsch}},  \emph {et~al.},\ }\bibfield  {title} {\enquote {\bibinfo {title} {Near edge x-ray absorption fine structure spectroscopy with x-ray free-electron lasers},}\ }\href@noop {} {\bibfield  {journal} {\bibinfo  {journal} {Appl. Phys. Lett.}\ }\textbf {\bibinfo {volume} {95}} (\bibinfo {year} {2009})}\BibitemShut {NoStop}%
\bibitem [{\citenamefont {Tanaka}\ \emph {et~al.}(2009)\citenamefont {Tanaka}, \citenamefont {Fukui}, \citenamefont {Hara}, \citenamefont {Hosoda}, \citenamefont {Inagaki}, \citenamefont {Inoue}, \citenamefont {Ishikawa}, \citenamefont {Kitamura}, \citenamefont {Hasegawa}, \citenamefont {Kano} \emph {et~al.}}]{tanaka2009high}%
  \BibitemOpen
  \bibfield  {author} {\bibinfo {author} {\bibfnamefont {Hitoshi}\ \bibnamefont {Tanaka}}, \bibinfo {author} {\bibfnamefont {T}~\bibnamefont {Fukui}}, \bibinfo {author} {\bibfnamefont {T}~\bibnamefont {Hara}}, \bibinfo {author} {\bibfnamefont {N}~\bibnamefont {Hosoda}}, \bibinfo {author} {\bibfnamefont {T}~\bibnamefont {Inagaki}}, \bibinfo {author} {\bibfnamefont {S}~\bibnamefont {Inoue}}, \bibinfo {author} {\bibfnamefont {T}~\bibnamefont {Ishikawa}}, \bibinfo {author} {\bibfnamefont {H}~\bibnamefont {Kitamura}}, \bibinfo {author} {\bibfnamefont {T}~\bibnamefont {Hasegawa}}, \bibinfo {author} {\bibfnamefont {Y}~\bibnamefont {Kano}},  \emph {et~al.},\ }\bibfield  {title} {\enquote {\bibinfo {title} {High performance sase fel achieved by stability-oriented accelerator system},}\ \ }(\bibinfo {year} {2009})\ pp.\ \bibinfo {pages} {758--765}\BibitemShut {NoStop}%
\bibitem [{\citenamefont {Shintake}\ \emph {et~al.}(2009)\citenamefont {Shintake}, \citenamefont {Tanaka}, \citenamefont {Hara}, \citenamefont {Tanaka}, \citenamefont {Togawa}, \citenamefont {Yabashi}, \citenamefont {Otake}, \citenamefont {Asano}, \citenamefont {Fukui}, \citenamefont {Hasegawa}, \citenamefont {Higashiya}, \citenamefont {Hosoda}, \citenamefont {Inagaki}, \citenamefont {Inoue}, \citenamefont {Kim}, \citenamefont {Kitamura}, \citenamefont {Kumagai}, \citenamefont {Maesaka}, \citenamefont {Matsui}, \citenamefont {Nagasono}, \citenamefont {Ohshima}, \citenamefont {Sakurai}, \citenamefont {Tamasaku}, \citenamefont {Tanaka}, \citenamefont {Tanikawa}, \citenamefont {Togashi}, \citenamefont {Wu}, \citenamefont {Kitamura}, \citenamefont {Ishikawa}, \citenamefont {Asaka}, \citenamefont {Bizen}, \citenamefont {Goto}, \citenamefont {Hirono}, \citenamefont {Ishii}, \citenamefont {Kimura}, \citenamefont {Kobayashi}, \citenamefont {Masuda}, \citenamefont {Matsushita}, \citenamefont {Mar\'echal},
  \citenamefont {Ohashi}, \citenamefont {Ohata}, \citenamefont {Shirasawa}, \citenamefont {Takagi}, \citenamefont {Takahashi}, \citenamefont {Takeuchi}, \citenamefont {Tanaka}, \citenamefont {Yamashita}, \citenamefont {Yanagida},\ and\ \citenamefont {Zhang}}]{shintake2009stable}%
  \BibitemOpen
  \bibfield  {author} {\bibinfo {author} {\bibfnamefont {Tsumoru}\ \bibnamefont {Shintake}}, \bibinfo {author} {\bibfnamefont {Hitoshi}\ \bibnamefont {Tanaka}}, \bibinfo {author} {\bibfnamefont {Toru}\ \bibnamefont {Hara}}, \bibinfo {author} {\bibfnamefont {Takashi}\ \bibnamefont {Tanaka}}, \bibinfo {author} {\bibfnamefont {Kazuaki}\ \bibnamefont {Togawa}}, \bibinfo {author} {\bibfnamefont {Makina}\ \bibnamefont {Yabashi}}, \bibinfo {author} {\bibfnamefont {Yuji}\ \bibnamefont {Otake}}, \bibinfo {author} {\bibfnamefont {Yoshihiro}\ \bibnamefont {Asano}}, \bibinfo {author} {\bibfnamefont {Toru}\ \bibnamefont {Fukui}}, \bibinfo {author} {\bibfnamefont {Teruaki}\ \bibnamefont {Hasegawa}}, \bibinfo {author} {\bibfnamefont {Atsushi}\ \bibnamefont {Higashiya}}, \bibinfo {author} {\bibfnamefont {Naoyasu}\ \bibnamefont {Hosoda}}, \bibinfo {author} {\bibfnamefont {Takahiro}\ \bibnamefont {Inagaki}}, \bibinfo {author} {\bibfnamefont {Shinobu}\ \bibnamefont {Inoue}}, \bibinfo {author} {\bibfnamefont {Yujong}\
  \bibnamefont {Kim}}, \bibinfo {author} {\bibfnamefont {Masanobu}\ \bibnamefont {Kitamura}}, \bibinfo {author} {\bibfnamefont {Noritaka}\ \bibnamefont {Kumagai}}, \bibinfo {author} {\bibfnamefont {Hirokazu}\ \bibnamefont {Maesaka}}, \bibinfo {author} {\bibfnamefont {Sakuo}\ \bibnamefont {Matsui}}, \bibinfo {author} {\bibfnamefont {Mitsuru}\ \bibnamefont {Nagasono}}, \bibinfo {author} {\bibfnamefont {Takashi}\ \bibnamefont {Ohshima}}, \bibinfo {author} {\bibfnamefont {Tatsuyuki}\ \bibnamefont {Sakurai}}, \bibinfo {author} {\bibfnamefont {Kenji}\ \bibnamefont {Tamasaku}}, \bibinfo {author} {\bibfnamefont {Yoshihito}\ \bibnamefont {Tanaka}}, \bibinfo {author} {\bibfnamefont {Takanori}\ \bibnamefont {Tanikawa}}, \bibinfo {author} {\bibfnamefont {Tadashi}\ \bibnamefont {Togashi}}, \bibinfo {author} {\bibfnamefont {Shukui}\ \bibnamefont {Wu}}, \bibinfo {author} {\bibfnamefont {Hideo}\ \bibnamefont {Kitamura}}, \bibinfo {author} {\bibfnamefont {Tetsuya}\ \bibnamefont {Ishikawa}}, \bibinfo {author} {\bibfnamefont
  {Takao}\ \bibnamefont {Asaka}}, \bibinfo {author} {\bibfnamefont {Teruhiko}\ \bibnamefont {Bizen}}, \bibinfo {author} {\bibfnamefont {Shunji}\ \bibnamefont {Goto}}, \bibinfo {author} {\bibfnamefont {Toko}\ \bibnamefont {Hirono}}, \bibinfo {author} {\bibfnamefont {Miho}\ \bibnamefont {Ishii}}, \bibinfo {author} {\bibfnamefont {Hiroaki}\ \bibnamefont {Kimura}}, \bibinfo {author} {\bibfnamefont {Toshiaki}\ \bibnamefont {Kobayashi}}, \bibinfo {author} {\bibfnamefont {Takemasa}\ \bibnamefont {Masuda}}, \bibinfo {author} {\bibfnamefont {Tomohiro}\ \bibnamefont {Matsushita}}, \bibinfo {author} {\bibfnamefont {Xavier}\ \bibnamefont {Mar\'echal}}, \bibinfo {author} {\bibfnamefont {Haruhiko}\ \bibnamefont {Ohashi}}, \bibinfo {author} {\bibfnamefont {Toru}\ \bibnamefont {Ohata}}, \bibinfo {author} {\bibfnamefont {Katsutoshi}\ \bibnamefont {Shirasawa}}, \bibinfo {author} {\bibfnamefont {Tetsuya}\ \bibnamefont {Takagi}}, \bibinfo {author} {\bibfnamefont {Sunao}\ \bibnamefont {Takahashi}}, \bibinfo {author}
  {\bibfnamefont {Masao}\ \bibnamefont {Takeuchi}}, \bibinfo {author} {\bibfnamefont {Ryotaro}\ \bibnamefont {Tanaka}}, \bibinfo {author} {\bibfnamefont {Akihiro}\ \bibnamefont {Yamashita}}, \bibinfo {author} {\bibfnamefont {Kenichi}\ \bibnamefont {Yanagida}}, \ and\ \bibinfo {author} {\bibfnamefont {Chao}\ \bibnamefont {Zhang}},\ }\bibfield  {title} {\enquote {\bibinfo {title} {Stable operation of a self-amplified spontaneous-emission free-electron laser in the extremely ultraviolet region},}\ }\href {\doibase 10.1103/PhysRevSTAB.12.070701} {\bibfield  {journal} {\bibinfo  {journal} {Phys. Rev. ST Accel. Beams}\ }\textbf {\bibinfo {volume} {12}},\ \bibinfo {pages} {070701} (\bibinfo {year} {2009})}\BibitemShut {NoStop}%
\bibitem [{\citenamefont {Saldin}\ \emph {et~al.}(2008)\citenamefont {Saldin}, \citenamefont {Schneidmiller},\ and\ \citenamefont {Yurkov}}]{saldin2008coherence}%
  \BibitemOpen
  \bibfield  {author} {\bibinfo {author} {\bibfnamefont {Evgeny~L}\ \bibnamefont {Saldin}}, \bibinfo {author} {\bibfnamefont {Evgeny~A}\ \bibnamefont {Schneidmiller}}, \ and\ \bibinfo {author} {\bibfnamefont {Mikhail~V}\ \bibnamefont {Yurkov}},\ }\bibfield  {title} {\enquote {\bibinfo {title} {Coherence properties of the radiation from x-ray free electron laser},}\ }\href@noop {} {\bibfield  {journal} {\bibinfo  {journal} {Opt. Commun.}\ }\textbf {\bibinfo {volume} {281}},\ \bibinfo {pages} {1179--1188} (\bibinfo {year} {2008})}\BibitemShut {NoStop}%
\bibitem [{\citenamefont {Thompson}(2017)}]{thompson2018possible}%
  \BibitemOpen
  \bibfield  {author} {\bibinfo {author} {\bibfnamefont {NR}~\bibnamefont {Thompson}},\ }\bibfield  {title} {\enquote {\bibinfo {title} {Possible method for the control of sase fluctuations},}\ }\href@noop {} {\bibfield  {journal} {\bibinfo  {journal} {in Proceedings of the 38th International Free Electron Laser Conference (FEL’17), Daejeon, Korea}\ }\textbf {\bibinfo {volume} {MOP039}},\ \bibinfo {pages} {129--131} (\bibinfo {year} {2017})}\BibitemShut {NoStop}%
\bibitem [{\citenamefont {Hara}\ \emph {et~al.}(2013)\citenamefont {Hara}, \citenamefont {Inubushi}, \citenamefont {Katayama}, \citenamefont {Sato}, \citenamefont {Tanaka}, \citenamefont {Tanaka}, \citenamefont {Togashi}, \citenamefont {Togawa}, \citenamefont {Tono}, \citenamefont {Yabashi} \emph {et~al.}}]{hara2013two}%
  \BibitemOpen
  \bibfield  {author} {\bibinfo {author} {\bibfnamefont {Toru}\ \bibnamefont {Hara}}, \bibinfo {author} {\bibfnamefont {Yuichi}\ \bibnamefont {Inubushi}}, \bibinfo {author} {\bibfnamefont {Tetsuo}\ \bibnamefont {Katayama}}, \bibinfo {author} {\bibfnamefont {Takahiro}\ \bibnamefont {Sato}}, \bibinfo {author} {\bibfnamefont {Hitoshi}\ \bibnamefont {Tanaka}}, \bibinfo {author} {\bibfnamefont {Takashi}\ \bibnamefont {Tanaka}}, \bibinfo {author} {\bibfnamefont {Tadashi}\ \bibnamefont {Togashi}}, \bibinfo {author} {\bibfnamefont {Kazuaki}\ \bibnamefont {Togawa}}, \bibinfo {author} {\bibfnamefont {Kensuke}\ \bibnamefont {Tono}}, \bibinfo {author} {\bibfnamefont {Makina}\ \bibnamefont {Yabashi}},  \emph {et~al.},\ }\bibfield  {title} {\enquote {\bibinfo {title} {Two-colour hard x-ray free-electron laser with wide tunability},}\ }\href@noop {} {\bibfield  {journal} {\bibinfo  {journal} {Nat. Commun.}\ }\textbf {\bibinfo {volume} {4}},\ \bibinfo {pages} {2919} (\bibinfo {year} {2013})}\BibitemShut {NoStop}%
\bibitem [{\citenamefont {Sun}\ \emph {et~al.}(2024)\citenamefont {Sun}, \citenamefont {Wang},\ and\ \citenamefont {Zhang}}]{Sun}%
  \BibitemOpen
  \bibfield  {author} {\bibinfo {author} {\bibfnamefont {Hao}\ \bibnamefont {Sun}}, \bibinfo {author} {\bibfnamefont {Xiaofan}\ \bibnamefont {Wang}}, \ and\ \bibinfo {author} {\bibfnamefont {Weiqing}\ \bibnamefont {Zhang}},\ }\bibfield  {title} {\enquote {\bibinfo {title} {Attosecond two-color x-ray free-electron lasers with dual chirp-taper configuration and bunching inheritance},}\ }\href {\doibase 10.1103/PhysRevAccelBeams.27.060701} {\bibfield  {journal} {\bibinfo  {journal} {Phys. Rev. Accel. Beams}\ }\textbf {\bibinfo {volume} {27}},\ \bibinfo {pages} {060701} (\bibinfo {year} {2024})}\BibitemShut {NoStop}%
\bibitem [{\citenamefont {Guo}\ \emph {et~al.}(2024)\citenamefont {Guo}, \citenamefont {Driver}, \citenamefont {Beauvarlet}, \citenamefont {Cesar}, \citenamefont {Duris}, \citenamefont {Franz}, \citenamefont {Alexander}, \citenamefont {Bohler}, \citenamefont {Bostedt}, \citenamefont {Averbukh} \emph {et~al.}}]{guo2024experimental}%
  \BibitemOpen
  \bibfield  {author} {\bibinfo {author} {\bibfnamefont {Zhaoheng}\ \bibnamefont {Guo}}, \bibinfo {author} {\bibfnamefont {Taran}\ \bibnamefont {Driver}}, \bibinfo {author} {\bibfnamefont {Sandra}\ \bibnamefont {Beauvarlet}}, \bibinfo {author} {\bibfnamefont {David}\ \bibnamefont {Cesar}}, \bibinfo {author} {\bibfnamefont {Joseph}\ \bibnamefont {Duris}}, \bibinfo {author} {\bibfnamefont {Paris~L}\ \bibnamefont {Franz}}, \bibinfo {author} {\bibfnamefont {Oliver}\ \bibnamefont {Alexander}}, \bibinfo {author} {\bibfnamefont {Dorian}\ \bibnamefont {Bohler}}, \bibinfo {author} {\bibfnamefont {Christoph}\ \bibnamefont {Bostedt}}, \bibinfo {author} {\bibfnamefont {Vitali}\ \bibnamefont {Averbukh}},  \emph {et~al.},\ }\bibfield  {title} {\enquote {\bibinfo {title} {Experimental demonstration of attosecond pump--probe spectroscopy with an x-ray free-electron laser},}\ }\href@noop {} {\bibfield  {journal} {\bibinfo  {journal} {Nat. Photonics}\ ,\ \bibinfo {pages} {1--7}} (\bibinfo {year} {2024})}\BibitemShut {NoStop}%
\bibitem [{\citenamefont {Bonifacio}\ \emph {et~al.}(1992)\citenamefont {Bonifacio}, \citenamefont {Corsini},\ and\ \citenamefont {Pierini}}]{bonifacio1992theory}%
  \BibitemOpen
  \bibfield  {author} {\bibinfo {author} {\bibfnamefont {R}~\bibnamefont {Bonifacio}}, \bibinfo {author} {\bibfnamefont {R}~\bibnamefont {Corsini}}, \ and\ \bibinfo {author} {\bibfnamefont {P}~\bibnamefont {Pierini}},\ }\bibfield  {title} {\enquote {\bibinfo {title} {Theory of the high-gain optical klystron},}\ }\href@noop {} {\bibfield  {journal} {\bibinfo  {journal} {Phys. Rev. A}\ }\textbf {\bibinfo {volume} {45}},\ \bibinfo {pages} {4091} (\bibinfo {year} {1992})}\BibitemShut {NoStop}%
\bibitem [{\citenamefont {Ding}\ \emph {et~al.}(2006)\citenamefont {Ding}, \citenamefont {Emma}, \citenamefont {Huang},\ and\ \citenamefont {Kumar}}]{ding2006optical}%
  \BibitemOpen
  \bibfield  {author} {\bibinfo {author} {\bibfnamefont {Yuantao}\ \bibnamefont {Ding}}, \bibinfo {author} {\bibfnamefont {Paul}\ \bibnamefont {Emma}}, \bibinfo {author} {\bibfnamefont {Zhirong}\ \bibnamefont {Huang}}, \ and\ \bibinfo {author} {\bibfnamefont {Vinit}\ \bibnamefont {Kumar}},\ }\bibfield  {title} {\enquote {\bibinfo {title} {Optical klystron enhancement to self-amplified spontaneous emission free electron lasers},}\ }\href@noop {} {\bibfield  {journal} {\bibinfo  {journal} {Phys. Rev. ST Accel. Beams}\ }\textbf {\bibinfo {volume} {9}},\ \bibinfo {pages} {070702} (\bibinfo {year} {2006})}\BibitemShut {NoStop}%
\bibitem [{\citenamefont {Geloni}\ \emph {et~al.}(2021)\citenamefont {Geloni}, \citenamefont {Guetg}, \citenamefont {Serkez},\ and\ \citenamefont {Schneidmiller}}]{geloni2021revision}%
  \BibitemOpen
  \bibfield  {author} {\bibinfo {author} {\bibfnamefont {Gianluca}\ \bibnamefont {Geloni}}, \bibinfo {author} {\bibfnamefont {Marc}\ \bibnamefont {Guetg}}, \bibinfo {author} {\bibfnamefont {Svitozar}\ \bibnamefont {Serkez}}, \ and\ \bibinfo {author} {\bibfnamefont {Evgeny}\ \bibnamefont {Schneidmiller}},\ }\bibfield  {title} {\enquote {\bibinfo {title} {Revision of optical klystron enhancement effects in self-amplified spontaneous emission free electron lasers},}\ }\href@noop {} {\bibfield  {journal} {\bibinfo  {journal} {Phys. Rev. Accel. Beams}\ }\textbf {\bibinfo {volume} {24}},\ \bibinfo {pages} {090702} (\bibinfo {year} {2021})}\BibitemShut {NoStop}%
\bibitem [{\citenamefont {Penco}\ \emph {et~al.}(2015)\citenamefont {Penco}, \citenamefont {Allaria}, \citenamefont {De~Ninno}, \citenamefont {Ferrari},\ and\ \citenamefont {Giannessi}}]{penco2015experimental}%
  \BibitemOpen
  \bibfield  {author} {\bibinfo {author} {\bibfnamefont {G}~\bibnamefont {Penco}}, \bibinfo {author} {\bibfnamefont {Enrico}\ \bibnamefont {Allaria}}, \bibinfo {author} {\bibfnamefont {Giovanni}\ \bibnamefont {De~Ninno}}, \bibinfo {author} {\bibfnamefont {Eugenio}\ \bibnamefont {Ferrari}}, \ and\ \bibinfo {author} {\bibfnamefont {L}~\bibnamefont {Giannessi}},\ }\bibfield  {title} {\enquote {\bibinfo {title} {Experimental demonstration of enhanced self-amplified spontaneous emission by an optical klystron},}\ }\href@noop {} {\bibfield  {journal} {\bibinfo  {journal} {Phys. Rev. Lett.}\ }\textbf {\bibinfo {volume} {114}},\ \bibinfo {pages} {013901} (\bibinfo {year} {2015})}\BibitemShut {NoStop}%
\bibitem [{\citenamefont {Penco}\ \emph {et~al.}(2017)\citenamefont {Penco}, \citenamefont {Allaria}, \citenamefont {De~Ninno}, \citenamefont {Ferrari}, \citenamefont {Giannessi}, \citenamefont {Roussel},\ and\ \citenamefont {Spampinati}}]{penco2017optical}%
  \BibitemOpen
  \bibfield  {author} {\bibinfo {author} {\bibfnamefont {Giuseppe}\ \bibnamefont {Penco}}, \bibinfo {author} {\bibfnamefont {Enrico}\ \bibnamefont {Allaria}}, \bibinfo {author} {\bibfnamefont {Giovanni}\ \bibnamefont {De~Ninno}}, \bibinfo {author} {\bibfnamefont {Eugenio}\ \bibnamefont {Ferrari}}, \bibinfo {author} {\bibfnamefont {Luca}\ \bibnamefont {Giannessi}}, \bibinfo {author} {\bibfnamefont {Eléonore}\ \bibnamefont {Roussel}}, \ and\ \bibinfo {author} {\bibfnamefont {Simone}\ \bibnamefont {Spampinati}},\ }\bibfield  {title} {\enquote {\bibinfo {title} {Optical klystron enhancement to self amplified spontaneous emission at fermi},}\ }\href@noop {} {\bibfield  {journal} {\bibinfo  {journal} {Photonics}\ }\textbf {\bibinfo {volume} {4}} (\bibinfo {year} {2017})}\BibitemShut {NoStop}%
\bibitem [{\citenamefont {Prat}\ \emph {et~al.}(2021)\citenamefont {Prat}, \citenamefont {Ferrari}, \citenamefont {Calvi}, \citenamefont {Ganter}, \citenamefont {Reiche},\ and\ \citenamefont {Schmidt}}]{prat2021demonstration}%
  \BibitemOpen
  \bibfield  {author} {\bibinfo {author} {\bibfnamefont {Eduard}\ \bibnamefont {Prat}}, \bibinfo {author} {\bibfnamefont {Eugenio}\ \bibnamefont {Ferrari}}, \bibinfo {author} {\bibfnamefont {Marco}\ \bibnamefont {Calvi}}, \bibinfo {author} {\bibfnamefont {Romain}\ \bibnamefont {Ganter}}, \bibinfo {author} {\bibfnamefont {Sven}\ \bibnamefont {Reiche}}, \ and\ \bibinfo {author} {\bibfnamefont {Thomas}\ \bibnamefont {Schmidt}},\ }\bibfield  {title} {\enquote {\bibinfo {title} {Demonstration of a compact x-ray free-electron laser using the optical klystron effect},}\ }\href@noop {} {\bibfield  {journal} {\bibinfo  {journal} {Appl. Phys. Lett.}\ }\textbf {\bibinfo {volume} {119}} (\bibinfo {year} {2021})}\BibitemShut {NoStop}%
\bibitem [{\citenamefont {Kittel}\ \emph {et~al.}(2024)\citenamefont {Kittel}, \citenamefont {Calvi}, \citenamefont {Reiche}, \citenamefont {Sammut}, \citenamefont {Wang},\ and\ \citenamefont {Prat}}]{kittel2024enhanced}%
  \BibitemOpen
  \bibfield  {author} {\bibinfo {author} {\bibfnamefont {Christoph}\ \bibnamefont {Kittel}}, \bibinfo {author} {\bibfnamefont {Marco}\ \bibnamefont {Calvi}}, \bibinfo {author} {\bibfnamefont {Sven}\ \bibnamefont {Reiche}}, \bibinfo {author} {\bibfnamefont {Nicholas}\ \bibnamefont {Sammut}}, \bibinfo {author} {\bibfnamefont {Guanglei}\ \bibnamefont {Wang}}, \ and\ \bibinfo {author} {\bibfnamefont {Eduard}\ \bibnamefont {Prat}},\ }\bibfield  {title} {\enquote {\bibinfo {title} {Enhanced x-ray free-electron laser performance with optical klystron and helical undulators},}\ }\href@noop {} {\bibfield  {journal} {\bibinfo  {journal} {J. Synchrotron Radiat.}\ } (\bibinfo {year} {2024})}\BibitemShut {NoStop}%
\bibitem [{\citenamefont {Reiche}(1999)}]{reiche1999genesis}%
  \BibitemOpen
  \bibfield  {author} {\bibinfo {author} {\bibfnamefont {Sven}\ \bibnamefont {Reiche}},\ }\bibfield  {title} {\enquote {\bibinfo {title} {Genesis 1.3: a fully 3d time-dependent fel simulation code},}\ }\href@noop {} {\bibfield  {journal} {\bibinfo  {journal} {Nucl. Instrum. Methods Phys. Res., Sect. A}\ }\textbf {\bibinfo {volume} {429}},\ \bibinfo {pages} {243--248} (\bibinfo {year} {1999})}\BibitemShut {NoStop}%
\bibitem [{\citenamefont {Prat}\ \emph {et~al.}(2016)\citenamefont {Prat}, \citenamefont {Calvi}, \citenamefont {Ganter}, \citenamefont {Reiche}, \citenamefont {Schietinger},\ and\ \citenamefont {Schmidt}}]{prat2016undulator}%
  \BibitemOpen
  \bibfield  {author} {\bibinfo {author} {\bibfnamefont {Eduard}\ \bibnamefont {Prat}}, \bibinfo {author} {\bibfnamefont {Marco}\ \bibnamefont {Calvi}}, \bibinfo {author} {\bibfnamefont {Romain}\ \bibnamefont {Ganter}}, \bibinfo {author} {\bibfnamefont {Sven}\ \bibnamefont {Reiche}}, \bibinfo {author} {\bibfnamefont {Thomas}\ \bibnamefont {Schietinger}}, \ and\ \bibinfo {author} {\bibfnamefont {Thomas}\ \bibnamefont {Schmidt}},\ }\bibfield  {title} {\enquote {\bibinfo {title} {Undulator beamline optimization with integrated chicanes for x-ray free-electron-laser facilities},}\ }\href@noop {} {\bibfield  {journal} {\bibinfo  {journal} {J. Synchrotron Radiat.}\ }\textbf {\bibinfo {volume} {23}},\ \bibinfo {pages} {861--868} (\bibinfo {year} {2016})}\BibitemShut {NoStop}%
\bibitem [{\citenamefont {Yu}(1991)}]{yu1991generation}%
  \BibitemOpen
  \bibfield  {author} {\bibinfo {author} {\bibfnamefont {Li~Hua}\ \bibnamefont {Yu}},\ }\bibfield  {title} {\enquote {\bibinfo {title} {Generation of intense uv radiation by subharmonically seeded single-pass free-electron lasers},}\ }\href@noop {} {\bibfield  {journal} {\bibinfo  {journal} {Phys. Rev. A}\ }\textbf {\bibinfo {volume} {44}},\ \bibinfo {pages} {5178} (\bibinfo {year} {1991})}\BibitemShut {NoStop}%
\end{thebibliography}%

\end{document}